\documentclass[a4paper,11pt]{article}
\usepackage{authblk}
\usepackage[skins]{tcolorbox}
\usepackage{mathtools,slashed}
\usepackage[T1]{fontenc}
\usepackage{slashed,verbatim,subfigure}
\usepackage[numbers,sort&compress]{natbib}
\usepackage{amsmath}
\usepackage{xcolor}
\usepackage{braket}
\usepackage{mathrsfs}
\usepackage{titlesec}
\usepackage{eucal}
\usepackage{amsbsy}
\usepackage[utf8]{inputenc}
\usepackage{amssymb}
\usepackage{mathtools}
\usepackage{appendix}
\usepackage{graphicx}

\usepackage[final]{pdfpages}
\usepackage{dcolumn}
\usepackage{bm}
\usepackage{upgreek}
\usepackage[a4paper]{geometry}
\usepackage{graphicx}
\usepackage{caption}
\usepackage{url}
\usepackage[colorlinks=true,
linkcolor=blue,    
urlcolor=green,   
citecolor=blue     
]{hyperref}

\def\gbm#1{{\let\alpha\upalpha \let\phi\upphi \let\lambda\uplambda \let\mu\upmu \let\rho\uprho \let\sigma\upsigma \let\tau\uptau \let\theta\uptheta \let\eta\upeta \let\xi\upxi \let\zeta\upzeta \bm{#1}}}

\catcode`@11
\def\seceqaa{\@addtoreset{equation}{section}
	\def\theequation{A\arabic{equation}}}
\def\seceqbb{\@addtoreset{equation}{section}
	\def\theequation{B\arabic{equation}}}
\def\seceqcc{\@addtoreset{equation}{section}
	\def\theequation{C\arabic{equation}}}
\def\seceqdd{\@addtoreset{equation}{section}
	\def\theequation{D\arabic{equation}}}
\def\seceqee{\@addtoreset{equation}{section}
	\def\theequation{E\arabic{equation}}}
\catcode`@11
\newcommand{\be}{\begin{eqnarray}}
	\newcommand{\ee}{\end{eqnarray}}
\usepackage{authblk}
\usepackage{bbm}
\newcommand*{\eins}{\ensuremath{\mathbbm 1}}
\begin{document}
	\title{The Dirac--Bergmann approach to optimal control theory}

\author[1]{Davit Aghamalyan\thanks{davagham@gmail.com}}
\author[2]{Aleek Maity\thanks{aleek@cmi.ac.in}}
\author[3]{Varun Narasimhachar\thanks{varun.achar@gmail.com}}
\author[4]{V V Sreedhar\thanks{sreedhar@cmi.ac.in}}
\affil[1]{Singapore University of Technology and Design, 8 Somapah Road, 487372 Singapore}
\affil[2,4]{Chennai Mathematical Institute, Plot H1, SIPCOT IT Park, Siruseri, Chennai 603103, India}
\affil[3]{Institute of High Performance Computing (IHPC), Agency for Science, Technology and Research (A*STAR), 1 Fusionopolis Way, \#16-16 Connexis, Republic of Singapore 138632}

\date{\today}

\maketitle
\titleformat{\section}
{\normalfont\Large\bfseries}{\thesection}{1em}{}

\begin{abstract}
	
	We present a novel framework for optimal control in both classical and quantum systems. Our approach leverages the Dirac--Bergmann algorithm: a systematic method for formulating and solving constrained dynamical systems. In contrast to the standard Pontryagin Principle, which is used in control theory, our approach bypasses the need to perform a variation to obtain the optimal solution. Instead, the Dirac--Bergmann algorithm generates the optimal solution automatically, through the closure of the Poission Bracket algebra of the full set of constraints and the Hamiltonian. The efficacy of our framework is demonstrated through two quintessential examples: (1) the classical brachistochrone problem and (2) the time-optimal control of a generic quantum system, relevant for quantum technological applications. In the latter example, both closed and open quantum systems are discussed.
\end{abstract}

\newpage

\section{Introduction}  
Applications where an agent with limited control on a system needs to direct its behavior to his/her benefit are ubiquitous.
Optimal control theory \cite{kirk2004optimal,berkovitz2013optimal,gamkrelidze2013principles,sethi2021optimal,ansel2024introduction} gives the necessary tools to steer a dynamical system to achieve a specified objective
under a given set of constraints. It plays a pivotal role in many real-world applications, ranging from aeronautical engineering 
to robotics and financial systems.  At its core, optimal control theory addresses problems where the evolution of a system is governed by dynamical (typically, differential) equations and subject to performance criteria, typically formulated as cost functionals to be minimized or maximized.

Recognized as a cornerstone of optimal control theory, the Pontryagin maximum principle (PMP) \cite{ansel2024introduction,d2021introduction,boscain2021introduction} specifies the necessary conditions that an admissible control must satisfy in order to be considered optimal. PMP introduces an auxiliary Hamiltonian (which is conventionally called the Pontryagin Hamiltonian) defined on an extended phase space that includes both the system state and an adjoint variable. This fundamental principle is based on the equivalence between any given constrained optimization problem and an unconstrained optimization problem in a higher-dimensional space. PMP states that an optimal control must extremize the Pontryagin Hamiltonian pointwise. The reformulated optimal control problem results in a boundary value problem that identifies candidate solutions for the original control problem.

The application of optimal control theory to quantum systems originated in the context of molecular dynamics, with foundational work by Peirce \textit{et al.}\ \cite{peirce1988optimal}, followed by extensive developments \cite{tannor1992time,somloi1993controlled,ramakrishna1995controllability,pahari2023stochastic}. 
In recent years, quantum optimal control has acquired importance within quantum science and technology, particularly in the domains of quantum information processing, quantum sensing, and quantum computation. Quantum systems are inherently delicate, high-dimensional, and often noisy, making their control especially challenging.  Most importantly, achieving high-fidelity operations, such as the precise implementation of quantum gates \cite{martinis2009energy,khaneja2005optimal}, quantum state transfer \cite{ying2016optimal,zhang2016optimal}, or noise suppression \cite{aroch2024mitigating,gorman2012overcoming}, requires robust and efficient control strategies.

In this paper, we present a general formalism for optimal control based on the Dirac--Bergmann algorithm \cite{dirac2013lectures,hanson1976constrained,sundermeyer1982constrained,henneaux1992quantization,brown2022singular}, originally developed to study Hamiltonian systems with constraints. The Dirac--Bergmann algorithm ensures that the relevant constraints are satisfied at all times during the dynamical evolution of the system. In contrast to PMP, our approach bypasses the need to obtain the optimal solution by performing a variation. It captures the relevant structure of PMP in other ways: adjoint variables emerge as canonical conjugates, the control Hamiltonian arises from the total Hamiltonian of the system, and the computation of variations (derivatives) is replaced by one of Poisson brackets. The extremality condition of PMP is reinterpreted as a constraint stabilization requirement, captured by the closure of the algebra generated by the iterated application of Poisson brackets.

We illustrate our formalism in two representative time-optimal problems: the classical brachistochrone, which seeks the path of shortest descent under gravity, and the problem of minimizing the time for effecting a desired state transformation on a quantum system under bounded Hamiltonians, also called the quantum brachistochrone problem \cite{carlini2005quantum,mostafazadeh2007quantum,carlini2008time,wang2015quantum,campaioli2019algorithm,lam2021demonstration,koike2022quantum}. Besides illustrating our general framework, the latter is also of independent interest for control problems arising in quantum technologies. 
For example, to mitigate decoherence, which is prevalent in noisy open quantum systems, one might wish to achieve the desired state transfer in the shortest possible time, thus minimizing the system's interaction with the environment \cite{aroch2024mitigating}.  

The paper is organized as follows: Section \ref{S2} reviews PMP and its role in optimal control theory. Section \ref{S3} reviews the Dirac–Bergmann algorithm. In Section \ref{S4} we use the Dirac-Bergmann algorithm to develop a general method for the optimal control of classical systems and apply this novel approach to the classical brachistochrone problem. In Section \ref{S5}, we extend our formalism to the optimal control of quantum systems and apply it to the quantum brachistochrone problem. We also show that the method can be useful to 
study the optimal control of open quantum systems. We summarize our results in section \ref{S6}. In section \ref{S7} we present an outlook for future directions.

\section{A brief review of the Pontryagin method for optimal control}
\label{S2}
In this review, we will follow the development presented in \cite{ansel2024introduction}, with minor notational modifications. The basic setting for an optimal control problem is a dynamical system whose state is defined by coordinates $\mathbf x\equiv\left[x_{a}\right]_{a=1\dots,n}$ and governed by some laws of motion
\begin{equation}\label{dyn_con}
	\dot{\mathbf x}(t)=\mathbf F\left[\mathbf x(t),\mathbf u(t),t\right],
\end{equation}
where $t$ is the time and an overhead dot denotes a total derivative with respect to $t$. Here, $\mathbf u\equiv\left[u_{i}\right]_{i=1\dots,m}$ are some parameters or variables that influence the system's dynamics e.g.\ time-dependent strengths of external fields. In the context of optimal control, these variables are the controls that an agent can use to manipulate the system.

The object of interest in an optimal control problem is the cost functional
\begin{equation}\label{cost}
	C\left[\mathbf{x}(\cdot),\mathbf{u}(\cdot),t_f\right] = G\left[\mathbf x\left(t_f\right),t_f\right]+\int_{0}^{t_f}F_0\left[\mathbf x(t),\mathbf u(t),t\right]~dt.
\end{equation}
Here, $G$ is the so-called \emph{terminal cost}, which depends only on the final time $t_f$ and the terminal configuration of the system; $F_0$ is the so-called \emph{running cost}, a time-dependent value that needs to be summed over the entire time interval. The problem then is to find the control $\mathbf u^{\star}(t)$ that minimizes the cost functional, subject to the dynamical constraints given by the equations of motion and an initial condition $\mathbf x(t=0)=\mathbf x_0$.

This is done by defining an ``action'' $S$ related to the corresponding cost functional
\begin{equation}\label{action}
	S\left[\mathbf{x}(\cdot),\mathbf{u}(\cdot),\gbm{\Lambda}(\cdot),t_f\right]=G\left[\mathbf x\left(t_f\right),t_f\right]+\int_{0}^{t_f}\left[F_0\left[\mathbf x(t),\mathbf u(t),t\right]+\Lambda_{a}(t)\left(\dot{x}_{a}(t)-F_{a}\left[\mathbf x(t),\mathbf u(t),t\right]\right)\right]dt,
\end{equation}
where the dynamical constraints have been incorporated into the control problem via the Lagrange multipliers $\gbm\Lambda(t)\equiv\left[\Lambda_{a}(t)\right]_{a=1,..,n}$, whose collection is customarily also interpreted as an ``adjoint state'' of the system. The optimal control problem is thus associated with the Lagrangian
\begin{equation}\label{main_Lag}
	L\left[\mathbf x(t),\dot{\mathbf x}(t),\mathbf u(t),\gbm\Lambda(t),t\right]=F_0\left[\mathbf x(t),\mathbf u(t),t\right]+\Lambda_{a}(t)\left(\dot{x}_{a}(t)-F_{a}\left[\mathbf x(t),\mathbf u(t),t\right]\right),
\end{equation}
together with the terminal cost term (whose effect is to impose boundary conditions, as we will see presently). Note that this is \emph{not} a Lagrangian associated with the bare dynamical laws---it is constructed to also capture the controls and the running cost. Importantly, it governs the dynamics not of the original $n$-dimensional physical system but of an augmented system whose configuration includes, in addition to the original system's coordinates $\mathbf x$, also the controls $\mathbf u$ and the adjoint state $\gbm\Lambda$. To be sure, these additional ``coordinates'' are special in that their time derivatives don't explicitly appear in the Lagrangian. With that caveat, we can proceed to apply the variational principle on the action in eq.\eqref{action} to derive the equations of motion:
\begin{alignat}{3}
	\label{EL1}
	\dot{x}_a(t) &= F_a(t);&&(\hbox{Euler--Lagrange equation for }\Lambda_a)\\
	\label{EL2}
	\dot{\Lambda}_a(t) &= \frac{\partial F_0}{\partial x_a(t)}-\Lambda_{b}(t)\frac{\partial F_b}{\partial x_a(t)};\;&&(\hbox{Euler--Lagrange equation for }x_a)\\
	\label{EL3}
	\Lambda_a\left(t_f\right)&=-\frac{\partial G}{\partial x_a\left(t_f\right)};&&(\hbox{boundary condition})\\
	\label{EL4}
	\frac{\partial F_0}{\partial u_i} &= \Lambda_{a}(t)\frac{\partial F_a}{\partial u_i(t)},&&(\hbox{Euler--Lagrange equation for }u_i)
\end{alignat}
where $a,b=1\dots,n$ and $i=1\dots,m$. The boundary condition \eqref{EL3} follows from the initial condition $\mathbf x(0)=\mathbf x_0$ by imposing stationarity of the action under variations of $\mathbf{x}\left(t_f\right)$. In principle, one can solve the equations (\ref{EL1}--\ref{EL4}) to find the optimal form of the control and the corresponding optimal trajectories subject to the appropriate boundary conditions.

One can effect a Legendre transformation to pass to the Hamiltonian formulation. Note that $\Lambda_a=\frac{\partial L}{\partial\dot{x}_a}$: that is, the $\gbm\Lambda$ turn out to be the conjugate momenta to the $\mathbf x$. The corresponding Hamiltonian is known as the Pontryagin Hamiltonian:
\begin{align}
	H_P\left[\mathbf x,\mathbf u,\gbm{\Lambda},t\right] := \Lambda_{a}  \dot{x}_a - L =\Lambda_{a} F_a - F_0.
\end{align} 

The optimal trajectories and the optimal control are given by Hamilton's equations
\begin{align}
	\label{H1}
	&\dot{\Lambda}_a = -\frac{\partial H_P}{\partial x_a}\\
	&\dot{x}_a = \frac{\partial H_P}{\partial \Lambda_a}\\
	\label{H3}
	&\frac{\partial H_P}{\partial u_i}=0.
\end{align}

Equations (\ref{H1}--\ref{H3}) are equivalent to the equations (\ref{EL1}--\ref{EL4}). They suffice in cases where the optimal control values lie in the interior of their associated domain; in this setting, the above method for solving for optimal control is called the weak Pontryagin principle. The general case, where the optimal controls may take values on the boundaries of their domain, call for the strong Pontryagin maximum principle. Our goal is to apply the Dirac--Bergmann algorithm, which is a dynamics-centred method. Therefore, we only consider cases where the optimal controls lie in the interior and, correspondingly, the associated optimization problems reduce to effective locally-generated dynamics in some augmented phase space. We will also eschew cases with terminal costs, restricting to cost functionals involving only running costs.

\section{A brief review of the Dirac--Bergmann method for constrained dynamical systems}\label{S3}
This overview of constrained dynamical systems draws from \cite{brown2022singular,dirac2013lectures,hanson1976constrained,henneaux1992quantization,sundermeyer1982constrained}. In this section, we will consider a generic system with an $N$-dimensional configuration space, with canonical coordinates and momenta $(\mathbf q,\mathbf p)\equiv\left(q_k,p_k\right)$. In a constrained dynamical system, not all the phase space variables are independent: they may be constrained by some mutual relations. Such constraints are naturally present in systems with a singular Lagrangian, but for some purposes (e.g.\ optimal control problems) one may also impose some by hand. In the latter case, one can modify the Lagrangian to incorporate the additional constraints through Lagrange multipliers, yielding a singular effective Lagrangian. The singularity refers to the fact that the determinant of the Hessian matrix
\[
W_{kl}:=\frac{\partial^{2}L}{\partial \dot{q}_{k}\partial \dot{q}_{l}}  =  \frac{\partial p_{k}}{\partial \dot{q}_{l}}
\]
vanishes:
\begin{equation}\label{hessian}
	\det W=0.
\end{equation}
As such, its rank $M<N$. From eq.(\ref{hessian}) we can solve for $M$ of the velocities 
\begin{equation}\label{qdota}
	\dot{q}^{r}=f^{r}(\mathbf q,p_{s},\dot{q}^{\rho})\qquad r, s =1,..,M \qquad \hbox{and} \qquad \rho =M+1,...,N
\end{equation}
where $(N-M)$ $\dot{q}$ are undetermined. The canonical momenta can be represented as
\begin{align}
	p_{k}&=\tilde{g}_{k}(\mathbf q,\dot{q}^{r},\dot{q}^{\rho})=\tilde{g}_{k}\Bigl(\mathbf q,f^{r}(\mathbf q,p_{s},\dot{q}^{\rho}),\dot{q}^{\rho}\Bigl)  =  g_{k}(\mathbf q,p_{s},\dot{q}^{\rho})   \quad k=1,2...N
\end{align}
out of which some momenta are independent of the $\dot{q}$, as one can not solve them via eq.(\ref{hessian})
\begin{equation}
	p_\mu =g_{\mu}(\mathbf q,p_{s})\qquad \mu =M+1,...,N
\end{equation}
the aforementioned $(N-M)$ conditions are the primary constraints
\begin{equation}\label{primary}
	\chi_\mu (\mathbf q,\mathbf p) \approx 0 ~~~~\mu=M+1,...,N
\end{equation}
where $\chi$ are the primary constraints, $\mu$ is the number of primary constraints.  The dynamics of the system unfold on the constrained subspace, not on the full phase space. In eq.(\ref{primary}) and the remainder of this paper, we will follow the convention in constrained dynamics literature of using the symbol `$\approx$' to denote so-called \emph{weak equality}, defined as equality not on the entire phase space but only on the constrained subspace.

The canonical Hamiltonian $H_c$ can be modified using Lagrange multipliers to take the primary constraints into account, resulting in the so-called \emph{primary Hamiltonian}
\begin{equation}\label{p_h}
	H_p = H_c + \lambda_\mu \chi_\mu \approx H_c.
\end{equation}
A necessary requirement on the constraints is that they must be preserved under evolution by the primary Hamiltonian:
\begin{equation}\label{consistency}
	\dot{\chi}_{\tau}=\Bigl\{\chi_{\tau},H_{p}\Bigl\} \approx \Bigl\{\chi_{\tau},H_{c}\Bigl\}+\lambda_{\mu}\Bigl\{\chi_{\tau},\chi_{\mu}\Bigl\} \overset{!}{\approx} 0.
\end{equation}
The exclamation mark here signifies that this equality is required. This may be accomplished in one of the following ways:
\begin{enumerate}
	\item The constraints are preserved unconditionally in time, i.e.\ the required equality simply holds naturally by virtue of the primary constraints having vanishing Poisson brackets with the Hamiltonian.
	\item\label{case3} The constraints can be preserved in time by assigning suitable values to the Lagrange multipliers $\lambda_\mu$. These values may depend on the coordinates.
	\item\label{case2} Neither of the above conditions hold. In this case, we are forced to add the required equality as a new condition among the phase space variables, known as a secondary constraint.
\end{enumerate}
In case \ref{case2}, the new set of constraints again needs to go through the consistency checks (\ref{consistency}), and thereby again the above three cases. One has to turn this crank until all the primary constraints and the newly generated secondary constraints are preserved over time. We will consider only finite-dimensional systems, so that this iterative procedure of generating newer elements in the algebra is guaranteed to terminate after finitely many steps. After all the constraints are thus ensured to be preserved in time, we call the final set of all the constraints $\mathcal{C}\equiv\left\{\chi_\tau\right\}_\tau$.

Now we categorize every constraint $\chi_\tau$ as first-class or second-class, according to the following:
\begin{alignat}{5}
	\label{f_c}
	\text{First-class:}\quad&&\Bigl\{\chi_\tau , \chi_\mu \Bigl\} &\approx 0\;~&&\forall  \chi_\mu\in \mathcal{C};\\
	\label{s_c}
	\text{Second-class:}\quad&&\Bigl\{\chi_\tau , \chi_\mu \Bigl\} &\neq 0\;~&&\text{for at least one }\chi_\mu\in \mathcal{C}.
\end{alignat}
Any phase-space observable $A(\mathbf q,\mathbf p)$ satisfying the condition stated in eq.(\ref{f_c}) is analogously said to be first-class. The first-class constraints generate gauge redundancies following from the fact that the Lagrange multipliers associated with them in the Hamiltonian remain undetermined. One needs so-called \emph{gauge-fixing conditions} to convert these first-class constraints into second-class and thereby fix the associated Lagrange multipliers through the consistency condition (\ref{consistency}). Thus, we could have more Lagrange multipliers than we started with in the primary Hamiltonian (\ref{p_h}). Let us denote by $\bar{\mathcal{C}}\equiv\left\{\zeta_\alpha\right\}_\alpha$ the final set of second-class constraints obtained by fixing all gauges in this manner. The set of all second-class constraints effectively reduces the physical phase space to a constrained subspace. We construct the following anti-symmetric matrix of the Poisson brackets amongst all the second-class constraints $\mathbf{\zeta}$:
\begin{equation}\label{cmn}
	D_{\alpha\beta} = \Bigl\{\zeta_\alpha ,\zeta_\beta\Bigl\}.
\end{equation}
Note that this matrix is guaranteed to be invertible, because all mutual dependencies have been eliminated by this point. We then define the effective Hamiltonian as
\begin{equation}\label{ef_h}
	H^\prime = H_c + \lambda_\alpha  \zeta_\alpha,
\end{equation}
where $\zeta_\alpha\in\bar{\mathcal{C}}$ are the final second-class constraints and $\lambda_\alpha$ the associated Lagrange multipliers.

These remaining constraints are preserved over time, since they account for all the primary and secondary constraints. Therefore,
\begin{align}\label{k1}
	\nonumber
	\dot{\zeta}_{\alpha}=\Bigl\{\zeta_{\alpha},H^{\prime}\Bigl\}= \Bigl\{\zeta_{\alpha},H_{c}\Bigl\}+\lambda_{\beta}\Bigl\{\zeta_{\alpha},\zeta_{\beta}\Bigl\} \approx 0\\ 
	\implies \lambda_\beta(\mathbf q,\mathbf p)= - \Bigl\{H_c ,\zeta_\alpha\Bigl\}D^{-1}_{\alpha\beta},
\end{align}
where we used the above-noted invertibility of $D$. Thereby, eq.(\ref{ef_h}) yields
\begin{equation}
	H^{\prime} = H_c - \zeta_\beta D^{-1}_{\alpha\beta} \Bigl\{H_c, \zeta_\alpha\Bigl\}
\end{equation}
Thus, the dynamics of any phase-space observable $A(\mathbf q,\mathbf p)$ is given by
\begin{align}\label{m_e}
	\dot{A} = \Bigl\{A, H^\prime \Bigl\} 
	= \Bigl\{A, H_c\Bigl\} - \Bigl\{A,\zeta_\alpha\Bigl\} D^{-1}_{\alpha\beta} \Bigl\{\zeta_\beta,H_c\Bigl\} 
	= \Bigl\{A, H_c\Bigl\}_{\mathrm{DB}}
\end{align}
where the last line employs the so-called \emph{Dirac bracket}, defined for any two phase space variables $A(\mathbf q,\mathbf p)$, $B(\mathbf q,\mathbf p)$ as
\begin{equation}
	\Bigr\{A , B\Bigr\}_{\mathrm{DB}} = \Bigr\{A ,B\Bigr\}-\Bigr\{A ,\zeta_\alpha\Bigr\} D^{-1}_{\alpha\beta}\Bigr\{\zeta_\beta, B\Bigr\}
\end{equation}
Now, consider the Dirac bracket of any phase space variable $A(\mathbf q,\mathbf p)$ with the second-class constraints
\begin{equation}
	\Bigr\{A, \zeta_\alpha\Bigr\}_{\mathrm{DB}} = \Bigr\{A, \zeta_\alpha\Bigr\}-\Bigr\{A, \zeta_\beta\Bigr\}D^{-1}_{\beta\gamma}\Bigr\{\zeta_\gamma, \zeta_\alpha\Bigr\}=0.
\end{equation}
Thus, by replacing all Poisson brackets with Dirac brackets, we can consistently impose the second-class constraints as strong equalities (i.e.,\ set them to zero identically) before evaluating the Dirac brackets themselves.

\section{The Dirac--Bergmann approach to optimal classical control}
\label{S4}
In this section, we outline our approach to optimal control problems, based on the Dirac--Bergmann algorithm. We begin by treating the original system with control parameters as a constrained dynamical system given by the Lagrangian ({\ref{main_Lag}}) of the optimal control theory
\begin{equation}
	L\left[\mathbf x(t),\dot{\mathbf x}(t),\mathbf u(t),\gbm\Lambda(t),t\right]=F_0\left[\mathbf x(t),\mathbf u(t),t\right]+\Lambda_{a}(t)\left[\dot{x}_{a}(t)-F_{a}\left[\mathbf x(t),\mathbf u(t),t\right]\right].
\end{equation}
Now, to do the constrained analysis, we promote the adjoint states $\mathbf{\Lambda}(t)$ and the control parameters $\mathbf{u}(t)$ to be on the same footing as the physical states $\mathbf{x}(t)$ by considering them to be generalized coordinates. The canonical conjugate momenta are given by
\begin{align}
	p_a &= \frac{\partial L}{\partial \dot{x}_a} = \Lambda_{a}; \\ 
	\pi_a &= \frac{\partial L}{\partial \dot{\Lambda}_a} = 0; \\
	\eta_i &= \frac{\partial L}{\partial \dot{u}_i} = 0.
\end{align}
which gives us the primary constraints
\begin{align}
	\label{PR1}
	\chi^{x}_{a} = p_a - \Lambda_{a} & \approx 0; \qquad a=1,..,n \\
	\chi^{\Lambda}_{a}=\pi_a  & \approx 0; \\
	\label{PR3}
	\chi^{u}_{i}=\eta_{i}  & \approx 0; \qquad i=1,..,m
\end{align}
where $m$ is the number of control parameters and $n$ is the number of system states and their adjoint states. 

The canonical Hamiltonian can be found by the Legendre transformation of $L$
\begin{equation}\label{CL_C}
	H_{c}(\mathbf x,\mathbf{\Lambda},\mathbf u,t) = \Lambda_{a} F_a  - F_0
\end{equation}

One can add the primary constraints to $H_c$, which gives us the primary Hamiltonian $H_p$. Thus 
\begin{equation}\label{QM_C}
	H_{p}(\mathbf x,\mathbf{\Lambda},\mathbf u,t) = H_c + \lambda_\mu \chi_\mu = \Lambda_{a} F_a - F_0 + \lambda^{x}_{a} (p_a - \Lambda_{a}) + \lambda^{\Lambda}_{a} \pi_a + \lambda^{u}_{i} \eta_i
\end{equation}
where $\chi_\mu$ are the primary constraints,  $\lambda_\mu$ are the associated Lagrange multipliers and $\mu=1,...,(2n+m)$ indexes the primary constraints, as can be seen from eqs.\ (\ref{PR1}--\ref{PR3}).

The primary constraints must be preserved in time. Thus we have
\begin{align}
	\label{l1}
	&\dot{\chi}^{x}_{a} =\bigr\{\chi^{x}_{a} ,H_p \bigr\} =  \bigr\{p_a , \Lambda_{b} F_b - F_0  \bigr\} - \lambda^{\Lambda}_{a} \neq 0 
	\implies  \lambda^{\Lambda}_{a} = \bigr\{p_a , \Lambda_{b} F_b - F_0  \bigr\}\\
	\label{l2}
	&\dot{\chi}^{\Lambda}_{a} =\bigr\{\pi_a ,H_p \bigr\} = F_a - \lambda^{x}_{a} \neq 0 \implies \lambda^{x}_{a} = F_a\\
	\label{chi_u}
	&\dot{\chi}^{u}_{i} =\bigr\{\eta_{i} ,H_p \bigr\} = \bigr\{\eta_{i} , \Lambda_a F_a -F_0 \bigr\} \neq 0
\end{align}
Note that all the Lagrange multipliers are fixed except for $\lambda^u_i$ for $i=1,...,m$. Consequently, eq.(\ref{chi_u}) generates $m$ secondary constraints. 

The rest of the Dirac--Bergmann algorithm proceeds as outlined in the previous section, although the details will depend on the nature of the problem. A spin-off of this exercise is that the optimal controls are automatically identified by the closure of the Poisson bracket algebra of the full set of constraints and the Hamiltonian, without the need to do a variation of the cost functional. Once the optimal controls are identified, it is straightforward to obtain the optimal trajectories by solving the corresponding equations of motion involving the Dirac bracket. The optimal control problem has thus been brought to a form suitable for the application of the Dirac--Bergmann algorithm, as advertised. 
The following table lists some salient, mutually-analogous features of the two approaches to a generic optimal control problem.

\noindent\resizebox{\textwidth}{!}{
\begin{tabular}{||c|c||}
 \hline
 \textbf{Pontryagin principle} & \textbf{Dirac--Bergmann algorithm}\\
 \hline\hline
 Variational approach & Algebraic approach\\
 \hline
 Lagrange multipliers treated as adjoint state variables 
 & Lagrange multipliers and their conjugate momenta 
 treated\\
 &on the same footing as other phase-space variables\\
\hline
 Pontryagin Hamiltonian & Effective Hamiltonian\\
\hline
 Derivatives w.r.t.\ control parameters & Closure of Poisson Bracket algebra\\
\hline
 Stationarity condition on Pontryagin Hamiltonian 
     &  Restricting to the physical subspace\\
 \hline
\end{tabular}
}

\subsection{An example: The classical brachistochrone}\label{S5}
The classical brachistochrone problem, originally posed by Bernoulli in 1696 \cite{Bernoulli1696}, concerns the trajectory
that minimizes the time taken for a point particle to fall from
an initial point $A$ to a final point $B$ (not exactly below it) 
in a uniform gravitational field.

This problem can be recast as a problem in optimal control theory by letting a time-dependent control force $\mathbf u$ guide the mass from point $A$ to point $B$ along a trajectory. The problem, then, is to find the optimal value of the control such that the mass reaches point $B$ in the minimum time possible. Once this is done, the PMP can be used to 
find the optimal trajectory \cite{djukic1976note}. In this section, we will use the Dirac--Bergmann approach to find the optimal control and trajectory for the same problem.

Considering the horizontal axis as the x-axis and the vertically downward axis as the y-axis, the equations of motion for the particle can be written as
\begin{align}\label{eom_cb}
	\ddot{x}=N\frac{\dot{y}}{v};~~~~\ddot{y}=g-N\frac{\dot{x}}{v}
\end{align}
where $g$ is the acceleration due to gravity, $N$ is the magnitude of the instantaneous normal force at any given point on the particle's trajectory, and $v$ is the speed along the tangent of the particle's trajectory.   

\begin{figure}[t]
    \centering
    \includegraphics[width=0.4\textwidth]{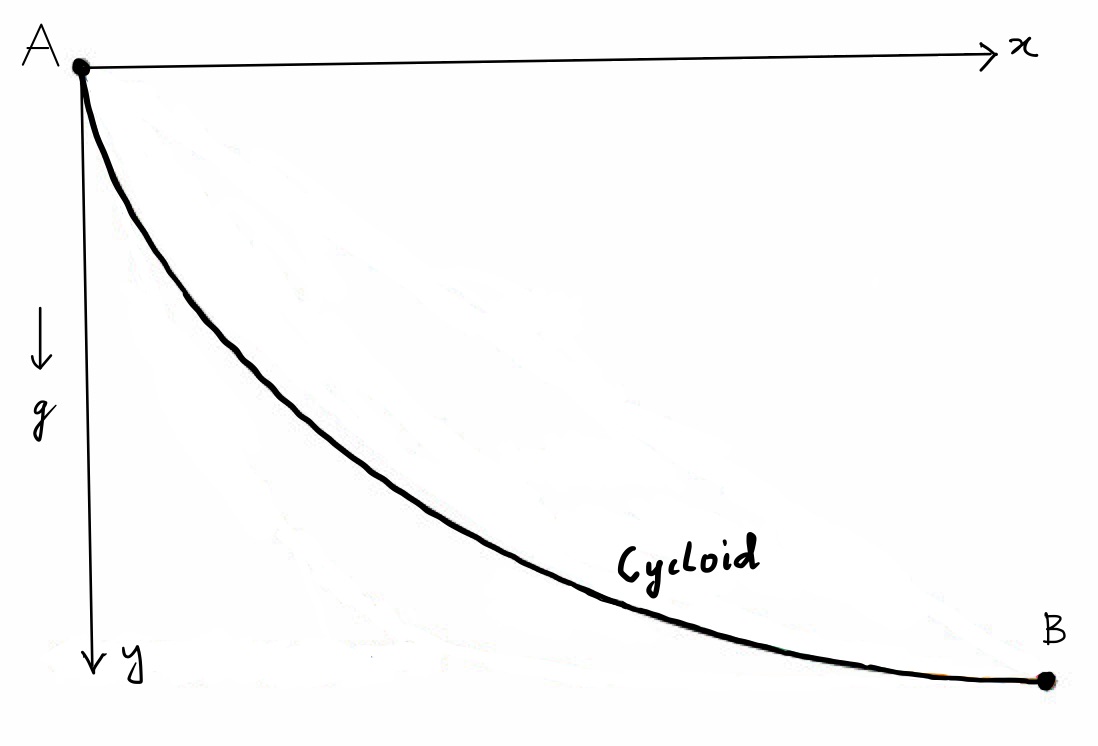}
    \caption{Classical Brachistochrone}
\end{figure}

To incorporate the dynamical constraints  (\ref{dyn_con}) we re-parametrize the equations of motion considering $x_1 = x,~ \dot{x_1}=x_2,~ x_3=y,~ \dot{x_3}=x_4$ to get
\begin{align}
	\dot{x}_1 = x_2; \qquad &F_1 = x_2\\
	\label{x2}
	\dot{x_2} = u; \qquad &F_2 = u\\
	\dot{x_3} = x_4; \qquad &F_3 = x_4\\
	\label{xy4}
	\dot{x_4} = g- u\frac{x_2}{x_4}; \qquad &F_4 = g- u\frac{x_2}{x_4}
\end{align}
where $u=\frac{N x_4}{v}$ is the control force.

The cost functional for optimal time control of the classical brachistochrone problem with $F_0 =1$ becomes
\begin{equation}
	C = \int_{0}^{t_f} F_0 ~dt = t_f  
\end{equation}
where $t_f$ is the final time.

The Lagrangian for the classical brachistochrone problem with the control force becomes
\begin{equation}
	L = F_0 + \mathbf\Lambda \cdot \left(\dot{\mathbf x} - \mathbf F\right) = 1 +\Lambda_1 \left(\dot{x_1}-x_2\right) + \Lambda_2 \left(\dot{x_2}-u\right) + \Lambda_3 \left(\dot{x_3}-x_4\right) + \Lambda_4 \left(\dot{x_4}-g+u\frac{x_2}{x_4}\right),
\end{equation}
where we omit the arguments of $L$ arguments for brevity. The set of canonical coordinates $x_a$, $\Lambda_a$ and $u$, and their conjugate momenta, are
\begin{align}
	p_a &= \frac{\partial L}{\partial \dot{x}_a} = \Lambda_a; \\
	\pi_a &= \frac{\partial L}{\partial \dot{\Lambda}_a} = 0; \\
	\eta &= \frac{\partial L}{\partial \dot{u}} = 0,
\end{align}
which gives the primary constraints of the form
\begin{align}
	\label{CC1}
	\chi^{x}_{a} = p_a - \Lambda_a &\approx 0;\\
	\label{CC2}
	\chi^{\Lambda}_{a} = \pi_{a} &\approx 0;\\
	\label{CC3}
	\chi^u = \eta &\approx 0. 
\end{align}
The canonical Hamiltonian is (\ref{CL_C})
\begin{equation}
	H_c = \Lambda_a F_a - 1  =-1 + \Lambda_1 x_2 +\Lambda_2 u +\Lambda_3 x_4 + \Lambda_4 \left(g- u \frac{x_2}{x_4}\right)
\end{equation}
and the primary Hamiltonian is (\ref{QM_C})
\begin{equation}
	\begin{split}
		H_p &= H_c +  \lambda^{x}_{a} \chi^{x}_{a} + \lambda^{\Lambda}_{a} \chi^{\Lambda}_{a} + \lambda^u \chi^u\\
		&=-1 + \Lambda_1 x_2 +\Lambda_2 u +\Lambda_3 x_4 + \Lambda_4 \left(g- u \frac{x_2}{x_4}\right) + \lambda^{x}_{a} \left(p_a - \Lambda_a\right) + \lambda^{\Lambda}_{a} \pi_a + \lambda^{u} \eta.
	\end{split}
\end{equation}
The consistency of the constraints eqs.\ (\ref{CC1}) and (\ref{CC2}) fix the Lagrange multipliers [eqs.\ (\ref{l1}) and (\ref{l2})] as follows:
\begin{align}
	\label{l_lambda}
	\lambda^{\Lambda}_{1} = 0,~~~\lambda^{\Lambda}_{2} = -\Lambda_1 + \Lambda_4 \frac{u}{x_4} ,~~~\lambda^{\Lambda}_{3} = 0,~~~\lambda^{\Lambda}_{4} = -\Lambda_3 - \Lambda_{4} u \frac{x_2}{x^2_4}, \\
	\label{l_x}
	\lambda^{x}_{1} = x_2 ,~~~\lambda^{x}_{2} = u ,~~~\lambda^{x}_{3} = x_4 ,~~~\lambda^{x}_{4} = g- u\frac{x_2}{x_{4}}. 
\end{align}
As noted in the previous section, following eq.(\ref{chi_u}), we see that all the Lagrange multipliers apart from $\lambda^u$ get fixed and we have a secondary constraint
\begin{equation}\label{CC4}
	\Sigma_1 = -\Lambda_2 + \Lambda_{4} \frac{x_2}{x_4}\approx 0.
\end{equation}
The consistency of $\Sigma_1$ gives another constraint
\begin{equation}\label{CC5}
	\Sigma_2 = \Lambda_{1} - \Lambda_3 \frac{x_2}{x_4} -g \Lambda_4 \frac{x_2}{x^{2}_{4}}\approx 0,
\end{equation}
while that of $\Sigma_2$ gives yet another constraint
\begin{equation}\label{CC6}
	\Sigma_3 = u \left(\frac{x^{2}_{4} + x^{2}_{2}}{x^{3}_{4}}\right) \left(-\Lambda_{3}-g\frac{\Lambda_{4}}{x_4}\right) + 2gx_2 \left(\frac{\Lambda_{3}x_4 + g \Lambda_{4}}{x^{3}_{4}}\right)\approx 0.
\end{equation}
The consistency of $\Sigma_3$ will fix the Lagrange multiplier $\lambda^u$ to be
\begin{equation}\label{uu}
	\lambda^u = \frac{-3 u^2 x_2^3 + 8 g u x_2^2 x_4 - 6 g^2 x_2 x_4^2 - 3 u^2 x_2 x_4^2 + 4 g u x_4^3}{x_4^2 \left(x_2^2 + x_4^2\right)}.
\end{equation}
Thus, there will be no further constraints that can be generated by this process.

All the primary and other constraints are given by eqs.\ (\ref{CC1}--\ref{CC3}) and (\ref{CC4}--\ref{CC6}). Note that all the constraints are second-class, since
\begin{align}
	&\Bigl\{\chi^{x}_{a}, \chi^{\Lambda}_{b}\Bigl\} = \Bigl\{p_a - \Lambda_a , \pi_b \Bigl\} = -\delta_{ab}\neq 0~~~ \hbox{for} ~~ a=b,~\hbox{where}~a,b = 1,2,3,4;\\
	&\Bigl\{\chi^u , \Sigma_3 \Bigl\} \neq 0;~~~~~
	\Bigl\{\Sigma_1 , \chi^{\Lambda}_{2} \Bigl\} \neq 0;~~~~~ \Bigl\{\Sigma_2 ,  \chi^{\Lambda}_{1}\Bigl\} \neq 0.
\end{align}
Now we relabel all the second-class constraints
\begin{alignat}{3}
	\label{sc1}
	\zeta_\alpha&=p_\alpha - \Lambda_\alpha \approx 0;&&(\alpha=1,2,3,4)\\
	\zeta_\beta&=\pi_\beta\approx 0;&&(\beta=5,6,7,8)\\
	\zeta_9&=\eta\approx 0;&&\\
	\zeta_{10}&=-\Lambda_2 + \Lambda_{4} \frac{x_2}{x_4}\approx 0;&&\\
	\zeta_{11}&=\Lambda_{1} - \Lambda_3 \frac{x_2}{x_4} -g \Lambda_4 \frac{x_2}{x^{2}_{4}}\approx 0;&&\\
	\label{CL12}
	\zeta_{12}&= u \left(\frac{x^{2}_{4} + x^{2}_{2}}{x^{3}_{4}}\right) \left(-\Lambda_{3}-g\frac{\Lambda_{4}}{x_4}\right) + 2gx_2 \left(\frac{\Lambda_{3}x_4 + g \Lambda_{4}}{x^{3}_{4}}\right)\approx 0.&&
\end{alignat}
The constraint that is of interest for the optimal control problem is $\zeta_{12}$ (\ref{CL12}) since it explicitly involves the control $u$. The optimal value of $u$ follows by setting $\zeta_{12}$ to be strongly zero, i.e.,\ restricting the system to the associated physical subspace. Thus we have
\begin{equation}\label{op_u}
	u^{\star} \left(\frac{x^{2}_{4} + x^{2}_{2}}{x^{3}_{4}}\right) \left(-\Lambda_{3}-g\frac{\Lambda_{4}}{x_4}\right) + 2gx_2 \left(\frac{\Lambda_{3}x_4 + g \Lambda_{4}}{x^{3}_{4}}\right)=0~\implies u^{\star} = \frac{2gx_2 x_4}{x^{2}_{2} + x^{2}_{4}}.
\end{equation}

It is important to emphasize that this is a generic feature: the second-class constraint that fixes $\lambda^u$ involves the control $u$ and thus dictates the optimal value of the control when the system is restricted to the corresponding physical subspace.

We conclude this section by using the Dirac--Bergmann approach to find the optimal trajectories corresponding to the optimal parameter $u^{\star}$. The equations of motion for the physical states $\mathbf x$ and the adjoint states $\mathbf{\Lambda}$ follow from the general
equation of motion \ (\ref{m_e}) and are given by
\begin{align}
	\label{x1}\dot{x}_1=x_2;~~~~~\qquad&\dot{\Lambda}_1 = 0; \\
	\dot{x}_2=\frac{2gx_2 x_4}{x^{2}_{2} + x^{2}_{4}};~~~~~\qquad&\dot{\Lambda}_2 = -\Lambda_1 + \Lambda_4 \frac{2gx_2 }{x^{2}_{2} + x^{2}_{4}};\\	\dot{x}_3=x_4;~~~~~\qquad&\dot{\Lambda}_3 = 0;\\
	\label{x4}
	\dot{x}_4=g\frac{x^2_4 - x^2_2}{x^{2}_{2} + x^{2}_{4}};~~~~~\qquad&\dot{\Lambda}_4 = -\Lambda_3-\Lambda_4  \frac{2gx^2_2}{x_4 (x^{2}_{2} + x^{2}_{4})},
\end{align}
where we have substituted the optimal value of $u$ from eq.(\ref{op_u}). In writing down the above equations, we need to use the matrix ${\mathbf D^{-1}}$ that appears in the Dirac bracket. 
This matrix is presented in the appendix. It is important to note that we have used the second-class constraint relations, eqs.\ (\ref{sc1}--\ref{CL12}), to rewrite the equations of motion in a smaller set of variables, as explained at the end of the last section. 

We can solve the equations for the physical states $\mathbf{x}$ [eqs.\ (\ref{x1}--\ref{x4})] to get
\begin{align}
	\label{xx1}
	x_{1} (t)&= \frac{c_1}{2}\left(t - \frac{c_1}{2g}\cos\theta\right) + c_3; \\
	x_{2}(t) &= \frac{c_1}{2}(1 - \sin\theta) =\dot{x_1}(t);\\
	x_3(t) &= c_4 - \frac{c_1^2}{4g}\sin\theta; \\
	\label{xx4}
	x_4(t) &= \frac{c_1}{2}\cos\theta =\dot{x_3}(t);\\
	\theta &= c_2 - \left(\frac{2g}{c_1}\right)t,
\end{align}
where $c_1, c_2, c_3, c_4$ are integration constants and $\theta$ is a phase parameter that traces the shape of the trajectory. Now with the initial boundary conditions $x_1(0)=0$, $x_3(0)=0$, $\dot{x_1}(0)=0$, and $\dot{x_3}(0)=0$, one can fix the integration constants to be
\begin{align}
	c_1 = \sqrt{4ga};~~~c_2=\frac{\pi}{2};~~~c_3=0;~~~c_4=a,  
\end{align}
where we have considered an overall horizontal shift by setting $c_3=0$ without changing the shape of the trajectory. Thus, eqs.\ (\ref{xx1}--\ref{xx4}) become
\begin{align}
	\label{cld1}x(t)=x_1(t)&=a\left(\frac{\pi}{2}-\theta-\cos\theta\right); \\
	\label{cld2}
	y(t)=x_3(t)&=a\left(1-\sin\theta\right).
\end{align}
Now we can do an overall phase shift $\theta^{\prime}= \frac{\pi}{2}-\theta$, again without changing the shape of the trajectory. Thus, eqs.\ (\ref{cld1}) and (\ref{cld2}) reduce to
\begin{align}
	x(t)&=a\left(\theta^\prime - \sin\theta^\prime\right);\\
	y(t)&=a\left(1-\cos\theta^\prime\right),  
\end{align}
which are readily recognized as the parametric equations of a cycloid, in agreement with \cite{djukic1976note}.


\section{Dirac--Bergmann approach to the optimal control of quantum systems}
In this section, we will apply the Dirac--Bergmann approach to the optimal control of quantum systems. As in the previous section, the relevant Lagrangian of the optimal control problem is
\begin{equation}
	\begin{split}
		\label{Lagrangian_Q}
		L &= F_0 (\ket{\psi},u, t) +  \bra{\phi}\Bigl({-i\ket{\dot{\psi}}}+ \hat{H}(t)\ket{\psi} \Bigl) + \Bigl({i\bra{\dot{\psi}}}+ \bra{\psi} \hat{H}(t) \Bigl)\ket{\phi}\\
		&=F_0 (\ket{\psi},u, t) -i\braket{\phi|\dot{\psi}} + \braket{\phi|F(\ket{\psi},u, t)} + i\braket{\dot{\psi}|\phi}+\braket{F(\ket{\psi},u, t)|\phi},
	\end{split}
\end{equation}
where the physical state of the system is $\ket{\psi}$, and $\bra{\phi}$ is its adjoint state, so that  $\ket{\psi}, \ket{\phi} \in \mathcal{H}$.  $\bra{\phi}$ is a Lagrange multiplier which has been introduced in the Lagrangian to enforce the dynamical constraint given by the Schr\"{o}dinger equation $i\ket{\dot{\psi}}=\hat{H}(t)\ket{\psi}$. $\hat{H}(t)$ is the Hamiltonian of the quantum 
system given by \cite{ansel2024introduction},
\begin{equation}
\hat{H}(t) = \hat{H_0} + \sum_{i=1}^{m} u_{i}(t) \hat{H}_{i}.
\end{equation}
The $u_i$ are the control parameters, $F_0$ is the running cost, and we define $\ket{F(\ket{\psi},u, t)} = \hat{H}(t)\ket{\psi}$. The last term in the Lagrangian is being added to make it a real quantity.

It should be noted that although the system being considered is 
quantum mechanical, the optimal control method being used is classical. This is similar to making measurements on a quantum system using 
a classical apparatus. With this caveat, we proceed to consider  $\ket{\psi}, \bra{\psi}, \ket{\phi}, \bra{\phi}, u_i$ formally as generalized coordinates. Their corresponding canonically conjugate momenta are thus 
 identified as 
\begin{equation}
	\begin{split}\label{p_q}
		\bra{P} = \frac{\partial L}{\partial\dot{\ket{\psi}}} = -i\bra{\phi}&;\qquad
		\ket{P} = \frac{\partial L}{\partial\dot{\bra{\psi}}} = i\ket{\phi}; \\
		\bra{\pi} = \frac{\partial L}{\partial \dot{\ket{\phi}}} = 0 &;\qquad
		\ket{\pi} = \frac{\partial L}{\partial\dot{\bra{\phi}}} = 0; \\
		&  \eta_i = \frac{\partial L}{\partial \dot{u}_i} = 0,
	\end{split}
\end{equation}
which allows us to postulate the following Poisson structure
\begin{align}
	\Bigl\{\ket{\psi},\bra{P}\Bigl\} = \eins;~~~\Bigl\{\bra{\psi},\ket{P}\Bigl\} = 1; \nonumber\\
	\Bigl\{\ket{\phi},\bra{\pi}\Bigl\} = \eins;~~~\Bigl\{\bra{\phi},\ket{\pi}\Bigl\} = 1;\\
	\Bigl\{u_i,\eta_i\Bigl\} = 1. \nonumber
\end{align}
The system is now primed for the application of the Dirac--Bergmann algorithm.

From eq.(\ref{p_q}) it follows that the primary constraints are 
\begin{align}
	\label{P1}
	\chi^{\ket{\psi}} = \bra{P} + i\bra{\phi} \approx 0; \\
	\label{P2}
	\chi^{\bra{\psi}} = \ket{P} - i\ket{\phi} \approx 0; \\
	\label{P3}
	\chi^{\ket{\pi}} = \bra{\pi} \approx 0;  \\
	\label{P4}
	\chi^{\bra{\pi}} = \ket{\pi} \approx 0; \\
	\label{P5}
	\chi^{u}_i = \eta_i \approx 0.
\end{align}
A Legendre transformation gives the canonical Hamiltonian for the control problem
\begin{equation}
	H_c = -F_0  - \braket{\phi|F} - \braket{F|\phi}.
\end{equation}
Incorporating the primary constraints, the canonical Hamiltonian becomes the primary Hamiltonian
\begin{equation}
	\begin{split}
		\label{Primary_Ham_Q}
		H_p  &= H_c + \lambda^{\ket{\psi}} \chi^{\ket{\psi}} + \lambda^{\bra{\psi}}\chi^{\bra{\psi}} + \lambda^{\ket{\phi}}\chi^{\ket{\phi}} + \lambda^{\bra{\phi}}\chi^{\bra{\phi}} + \lambda^{u}_i\chi^{u}_i\\
		&=-F_0  - \braket{\phi|F} - \braket{F|\phi} +
		\lambda^{\ket{\psi}} (\bra{P} + i\bra{\phi}) +
		\lambda^{\bra{\psi}} (\ket{P} - i\ket{\phi}) + \lambda^{\ket{\phi}} (\bra{\pi}) + \lambda^{\bra{\phi}} (\ket{\pi}) + \lambda^{u}_i \eta_i.
	\end{split}
\end{equation}
where $\lambda^{\ket{\psi}}$, $\lambda^{\bra{\psi}}$, $\lambda^{\ket{\phi}}$, $\lambda^{\bra{\phi}}$, $\lambda^u$ are other Lagrange multipliers to incorporate the primary constraints  $\chi^{\ket{\psi}}$, $\chi^{\bra{\psi}}$, $\chi^{\ket{\phi}}$, $\chi^{\bra{\phi}}$, $\chi^u$ in to the primary Hamiltonian respectively.

Requiring the primary constraints eqs.\ (\ref{P1}--\ref{P5}) to be preserved in time, we get
\begin{equation}
	\begin{split}\label{L1}
		\dot{\chi}^{\bra{\psi}} = \bigl\{\bra{P}+i\bra{\phi}, H_p \bigr\} &= \bigl\{\bra{P},-F_0 -\braket{\phi|F}-\braket{F|\phi} \bigr\} +i \lambda^{\bra{\phi}} \neq 0 \\
		&\implies \lambda^{\bra{\phi}}  = i\bigl\{\bra{P},F_0 - \braket{\phi | F}-\braket{F|\phi} \bigr\};
	\end{split}
\end{equation}
\begin{equation}\label{L2}
	\begin{split}
		\dot{\chi}^{\ket{\psi}} = \bigr\{\ket{P}-i\ket{\phi},H_p\bigr\} &=\bigr\{\ket{P}-i\ket{\phi},-F_0 -\braket{\phi|F}-\braket{F|\phi}  \bigr\} - i\lambda^{\ket{\phi}}\neq 0\\
		&\implies  \lambda^{\ket{\phi}} = -i \bigl\{\ket{P},F_0 - \braket{\phi | F}-\braket{F|\phi} \bigr\};
	\end{split}
\end{equation}
\begin{equation}\label{L3}
	\begin{split}
		\dot{\chi}^{\ket{\phi})} = \bigl\{\bra{\pi}), H_p \bigl\} &= \bra{F}+i\lambda^{\bra{\psi}} \neq 0 \\
		& \implies \lambda^{\bra{\psi}} = i\bra{F};
	\end{split}
\end{equation}
\begin{equation}\label{L4}
	\begin{split}
		\dot{\chi}^{\bra{\phi})} = \bigl\{\ket{\pi}), H_p \bigl\} &= \ket{F}-i\lambda^{\ket{\psi}} \neq 0 \\
		& \implies \lambda^{\ket{\psi}} =- i\ket{F};
	\end{split}
\end{equation}
\begin{equation}
	\begin{split}\label{q_chi_u}
		\dot{\chi}^{u}_i = \bigl\{\eta_i , H_p\} = \bigl\{\eta_i , -F_0 - \braket{\phi|F} - \braket{F|\phi} \} \neq 0.
	\end{split}
\end{equation}
From eqs.\ (\ref{L1}--\ref{q_chi_u}), we see that all the Lagrange multipliers are fixed except for $\lambda^u_i ,~i=1,...,m$. As in the case of the classical brachistochrone, eq.(\ref{q_chi_u}) can't be solved for Lagrange multipliers and produces secondary constraints. 

As explained in the last section, all we need to do now is to complete 
the Dirac--Bergmann analysis and identify the constraint that produces
the optimal value for the control parameter, whence the optimal 
trajectories can also be obtained in a straightforward manner.


\subsection{An example: the quantum brachistochrone}
In this section, we will apply the general method developed in the previous section to the quantum brachistrochrone problem. It is important to note that this problem is not merely a quantum version of the classical one of a particle falling under a gravitational field; rather, it is a very general class of problems, the defining feature being that the cost function to be optimized (specifically, minimized) is evolution time. The Lagrangian for the quantum brachistochrone problem is \cite{carlini2005quantum,carlini2008time,koike2022quantum} \footnote{Our Lagrangian differs from those of \cite{carlini2005quantum,koike2022quantum}, in which the authors use the Fubini--Study 
metric to define the infinitesimal line element $ds$. However, since the metric only determines how the state changes on the projective Hilbert space related to the distance element, but does not determine the dynamics itself, we can take it to be uniformly equal to unity without affecting the optimal trajectory and controls. For a related discussion, see \cite{carlini2008time}.}
\begin{equation}\label{Q_L}
	L = 1 + \Lambda\biggl(\frac{\mathrm{Tr}\tilde{H}^2}{2}-\omega^2 \biggl) - i \braket{\phi|\dot{\psi}} + \braket{\phi|H|\psi} + i \braket{\dot{\psi}|\phi} +\braket{\psi|H|\phi},
\end{equation}
where $\Lambda$ is a constant Lagrange multiplier that enforces the energy upper bound of the control via the constraint $\frac{\mathrm{Tr}\tilde{H}^2}{2}-\omega^2$, and $\bra{\phi}$ is a dynamical Lagrange multiplier which enforces the dynamical constraint coming from the Schr\"{o}dinger's equation, $i\ket{\dot{\psi}}=H\ket{\psi}$. The operator $\tilde{H}=H-\frac{\mathrm{Tr}H}{n}$ is the traceless part of the Hamiltonian $H$.

The canonical momenta following from the Lagrangian (\ref{Q_L}) are
\begin{align}
	\bra{P} = \frac{\partial L}{\partial\dot{\ket{\psi}}} = -i\bra{\phi}&;\qquad
	\ket{P} = \frac{\partial L}{\partial\dot{\bra{\psi}}} = i\ket{\phi}; \\
	\bra{\pi} = \frac{\partial L}{\partial \dot{\ket{\phi}}} = 0 &;\qquad
	\ket{\pi} = \frac{\partial L}{\partial\dot{\bra{\phi}}} = 0; \\
	P_H = \frac{\partial L}{\partial\ket{\dot{H}}} = 0;  ~~~~~~ &P_\Lambda = \frac{\partial L}{\partial\ket{\dot{\Lambda}}} = 0.
\end{align}

Thus, we have the primary constraints
\begin{align}
	\label{qp1}
	\chi^{\ket{\psi}} = \bra{P} + i\bra{\phi}\approx 0;~~~ &\chi^{\bra{\psi}} = \ket{P} - i\ket{\phi}\approx 0; \\
	\chi^{\ket{\phi}} = \bra{\pi}\approx 0;~~~ &\chi^{\bra{\phi}} = \ket{\pi} \approx 0;\\
	\label{qp3}
	\chi^{H} = P_H \approx 0 ;~~~ &\chi^{\Lambda} = P_\Lambda \approx 0.
\end{align}
The canonical Hamiltonian can be obtained by the Legendre transformation
\begin{equation}
	\begin{split}
		H_c =  \ket{\dot{\psi}} \bra{P}+ \ket{P} \bra{\dot{\psi}} +  \ket{\dot{\phi}}\bra{\pi}+\ket{\pi} \bra{\dot{\phi}} + P_H \dot{H} +P_\Lambda \dot{\Lambda} - L\\
		=-1-\Lambda \biggl(\frac{Tr\tilde{H}^2}{2}-\omega^2 \biggl) -\braket{\phi|H|\psi}-\braket{\psi|H|\phi}.
	\end{split}
\end{equation}
Similarly, the primary Hamiltonian is given by
\begin{equation}
	\begin{split}
		H_p 
		=-1-\Lambda \biggl(\frac{Tr\tilde{H}^2}{2}-\omega^2 \biggl) -\braket{\phi|H|\psi}-\braket{\psi|H|\phi} &+ \lambda^{\ket{\psi}}\Bigl(\bra{P}+i\bra{\phi}\Bigl)+\lambda^{\bra{\psi}}\Bigl(\ket{P}-i\ket{\phi}\Bigl)\\
		&+\lambda^{\ket{\phi}} \bra{\pi} +\lambda^{\bra{\phi}} \ket{\pi}+\lambda^{H} P_H + \lambda^{\Lambda} P_\Lambda.
	\end{split}
\end{equation}
The consistency of the primary constraints over time yields
\begin{align}
	\label{z1}
	&\dot{\chi}^{\ket{\psi}} = \Bigl\{\chi^{\ket{\psi}}  , H_p \Bigl\} = \bra{\phi}H  + i \lambda^{\bra{\phi}}\neq 0 \implies \lambda^{\bra{\phi}} = i \bra{\phi}H; \\
	\label{z2}
	&\dot{\chi}^{\bra{\psi}} = \Bigl\{\chi^{\bra{\psi}}  , H_p \Bigl\} =  H\ket{\phi} - i \lambda^{\ket{\phi}}\neq 0 \implies \lambda^{\ket{\phi}} = -iH\ket{\phi} ;\\
	\label{z3}
	&\dot{\chi}^{\ket{\phi}} = \Bigl\{\chi^{\ket{\phi}} , H_p \Bigl\}= \bra{\psi}H+i \lambda^{\bra{\psi}} \neq 0 \implies \lambda^{\bra{\psi}} = i \bra{\psi}H;\\
	\label{z4}
	&\dot{\chi}^{\bra{\phi}} = \Bigl\{\chi^{\bra{\phi}} , H_p \Bigl\}= H\ket{\psi}-i \lambda^{\ket{\psi}} \neq 0 \implies \lambda^{\ket{\psi}} = -i H\ket{\psi};\\
	\label{z5}
	&\dot{\chi}^{H} = \Bigl\{P_H , H_p \Bigl\} = \ket{\psi}\bra{\phi}+\ket{\phi}\bra{\psi} + \Lambda\tilde{H} \neq 0;\\
	\label{z6}
	&\dot{\chi}^{\Lambda} = \Bigl\{P_{\Lambda}, H_p\Bigl\} = \frac{\mathrm{Tr}\tilde{H}^2}{2}-\omega^2 \neq 0.
\end{align}
Eqs.\ (\ref{z5}) and (\ref{z6}) give the following secondary constraints:
\begin{align}
	\label{qs1}
	&\chi^{\Lambda^{\prime}} = \frac{\mathrm{Tr}\tilde{H}^2}{2}-\omega^2  \approx 0; \\
	\label{qs2}
	&\chi^{H^\prime} = \ket{\psi}\bra{\phi}+\ket{\phi}\bra{\psi} + \Lambda\tilde{H} \approx 0,
\end{align}
whose consistency entails
\begin{align}
	&\dot{\chi}^{\Lambda^{\prime}}=\Bigl\{\chi^{\Lambda^{\prime}}, H_p\Bigl\}= \lambda^H \tilde{H} \neq 0 \implies \lambda^H = 0, ~~~\hbox{since} ~~\tilde{H}\neq 0;\\
	&\dot{\chi}^{H^\prime} =\Bigl\{\chi^{H^\prime}, H_p\Bigl\}= \lambda^H \Lambda (1-\frac{1}{n}) + \lambda^{\Lambda} \tilde{H} \neq 0 \implies \lambda^\Lambda = 0,~~~\hbox{since}~~ \lambda^H = 0.
\end{align}
Thus, all the new Lagrange multipliers are fixed with all the constraints being preserved over time. 

The Poisson brackets between the constraints (both primary and secondary) follow from eqs.\ (\ref{qp1}--\ref{qp3}) and (\ref{qs1}--\ref{qs2}):
\begin{equation}
	\begin{split}
		&\Bigl\{\chi^{\ket{\psi}}, \chi^{H^{\prime}}\Bigl\} = -\bra{\phi};~~~ \Bigl\{\chi^{\bra{\psi}}, \chi^{H^{\prime}}\Bigl\} = -\ket{\phi}; \\
		&\Bigl\{\chi^{\ket{\phi}},\chi^{H^{\prime}}\Bigl\} = -\bra{\psi};~~~\Bigl\{\chi^{\bra{\phi}}, \chi^{H^{\prime}}\Bigl\} = -\ket{\psi};\\
		&\Bigl\{\chi^{\Lambda^{\prime}}, \chi^{H}\Bigl\} = \tilde{H}; ~~~ ~~~~~\Bigl\{\chi^{\Lambda}, \chi^{H^{\prime}}\Bigl\} = -\tilde{H}.
	\end{split}
\end{equation}
Thus, all the constraints are second-class. We relabel them as
\begin{align}
	\label{zeta1}
	&\zeta_1 = \bra{P} + i\bra{\phi}\approx 0;\\
	&\zeta_2 = \ket{P} - i\ket{\phi}\approx 0;\\
	&\zeta_3 = \bra{\pi}\approx 0;\\
	&\zeta_4 = \ket{\pi}\approx 0;\\
	&\zeta_5 = P_H\approx 0;\\
	&\zeta_6 = P_\Lambda\approx 0;\\
	\label{zeta7}
	&\zeta_7 =  \ket{\psi}\bra{\phi}+\ket{\phi}\bra{\psi} + \Lambda\tilde{H} \approx 0;\\
	\label{zeta8}
	&\zeta_8 = \frac{\mathrm{Tr}\tilde{H}^2}{2}-\omega^2  \approx 0.
\end{align}
The matrix of the Poisson brackets amongst all the second-class constraints is given by
\[ D_{\alpha \beta} =
\begin{pmatrix}
	0 &0 &0 &i &0 &0 &-\bra{\phi}  &0 \\
	0 &0 &-i &0 &0 &0 &-\ket{\phi}  &0 \\
	0 &i &0 &0 &0 &0 &-\bra{\psi}  &0 \\
	-i &0 &0 &0 &0 &0 &-\ket{\psi}  &0 \\
	0 &0 &0 &0 &0 &0 &-\Lambda(1-\frac{1}{n})  &-\tilde{H} \\
	0 &0 &0 &0 &0 &0 &-\tilde{H}  &0 \\
	\bra{\phi} &\ket{\phi} &\bra{\psi} &\ket{\psi} &\Lambda(1-\frac{1}{n}) &\tilde{H} &0  &0 \\
	0 &0 &0 &0 &\tilde{H} &0 &0  &0 \\
\end{pmatrix},
\]
and its inverse matrix

\[ D_{\alpha \beta}^{-1} =
\begin{pmatrix}
	0 &0 &0 &-\frac{1}{i} &0 &\frac{\ket{\psi}}{i\tilde{H}} &0  &0 \\
	0 &0 &\frac{1}{i} &0 &0 &-\frac{\bra{\psi}}{i\tilde{H}} &0  &0 \\
	0 &-\frac{1}{i} &0 &0 &0 &\frac{\ket{\phi}}{i\tilde{H}} &0  &0 \\
	\frac{1}{i} &0 &0 &0 &0 &-\frac{\bra{\phi}}{i\tilde{H}} &0  &0 \\
	0 &0 &0 &0 &0 &0 &0  &\frac{1}{\tilde{H}} \\
	-\frac{\ket{\psi}}{i\tilde{H}} &\frac{\bra{\psi}}{i\tilde{H}} &-\frac{\ket{\phi}}{i\tilde{H}} &\frac{\bra{\phi}}{i\tilde{H}} &0 &0 &\frac{1}{\tilde{H}}  &-\frac{\Lambda(1-\frac{1}{n})}{\tilde{H}^{2}} \\
	0 &0 &0 &0 &0 &-\frac{1}{\tilde{H}} &0  &0 \\
	0 &0 &0 &0 &-\frac{1}{\tilde{H}} &\frac{\Lambda(1-\frac{1}{n})}{\tilde{H}^{2}} &0  &0 \\
\end{pmatrix}.
\]
Now, on the constrained subspace, the conditions eqs.\ (\ref{zeta1}--\ref{zeta8}) hold strongly, i.e., 
\begin{align}
	\label{zee7}
	&\zeta_7 =  \ket{\psi}\bra{\phi}+\ket{\phi}\bra{\psi} + \Lambda\tilde{H} = 0;\\
	\label{zee8}
	&\zeta_8 = \frac{\mathrm{Tr}\tilde{H}^2}{2}-\omega^2  = 0.
\end{align}
The equations of motion for $\ket{\psi}$ and $\ket{\phi}$ can then be obtained by 
\begin{equation}
	\begin{split}
		\label{psi}
		\ket{\dot{\psi}} &= \bigl\{\ket{\psi}, H_c \bigl\} - \bigl\{\ket{\psi},\zeta_{\alpha} \bigl\} D^{-1}_{\alpha\beta} \bigl\{\zeta_{\beta},H_c \bigl\}
		= \frac{H\ket{\psi}}{i}  - \frac{\ket{\psi}}{i\tilde{H}} \biggl(\frac{\mathrm{Tr}\tilde{H}^2}{2}-\omega^2\biggl) \\
		&= \frac{H\ket{\psi}}{i};
	\end{split}
\end{equation}
\begin{equation}
	\label{phi}
	\begin{split}
		\ket{\dot{\phi}} &= \bigl\{\ket{\phi}, H_c \bigl\} - \bigl\{\ket{\phi},\zeta_{\alpha} \bigl\} D^{-1}_{\alpha\beta} \bigl\{\zeta_{\beta},H_c \bigl\}
		= \frac{H\ket{\phi}}{i}  - \frac{\ket{\phi}}{i\tilde{H}} \biggl(\frac{\mathrm{Tr}\tilde{H}^2}{2}-\omega^2\biggl) \\
		&= \frac{H\ket{\phi}}{i},
	\end{split}
\end{equation}
where in the last step of eqs.\ (\ref{psi}) and (\ref{phi}), we eliminate the last term using the second class constraint \ (\ref{zee8}). These equations of motion are seen to be in agreement with \cite{koike2022quantum}.

Now we can use eq.(\ref{zee7}) to find the optimal value of the control with the help of eqs.\ (\ref{zee8}--\ref{phi}). Taking the trace of eq.(\ref{zee7}),
\begin{equation}
	\label{im}
	\braket{\psi|\phi}^{\star} = -\braket{\psi|\phi}, \hbox{ since Tr} \tilde{H} = 0.
\end{equation} 
The expectation value of eq.({\ref{zee7}}) with respect to $\ket{\psi}$ then implies that
\begin{equation}
	\label{avg_H}
	\braket{\tilde{H}} = 0,~\hbox{which is equivalent to } \braket{H} = \frac{\mathrm{Tr}H}{n}.
\end{equation}
Applying eq.(\ref{zee7}) to $\ket{\psi}$,
\begin{equation}
	\label{fi}
	\ket{\phi} = \biggl(-\lambda \tilde{H} + \braket{\psi|\phi}\biggl)\ket{\psi}.
\end{equation}
Inserting this back into eq.(\ref{zee7}), we get
\begin{equation}
	\label{H_OP}
	\tilde{H} = \tilde{H} P + P \tilde{H},
\end{equation}
which gives the optimal Hamiltonian and $P=\Ket{\psi}\bra{\psi}$. From eqs.\ (\ref{avg_H}) and (\ref{H_OP}), we get
\begin{equation}\label{const}
	(\Delta E)^2 = \braket{H^2} - \braket{H}^2 = \braket{\tilde{H}^2} = \frac{\mathrm{Tr}\tilde{H}^2}{2} = \omega^2.
\end{equation}
Substituting eq.(\ref{fi}) back into eq.(\ref{zee7}), we get
\begin{equation}\label{OP}
	\frac{d}{dt} (\lambda \tilde{H})\ket{\psi}=0.
\end{equation}
Multiplying throughout by $\bra{\psi}H$, we get $\lambda$ to be a constant. Subsequently, eq.(\ref{OP}) implies 
\begin{equation}\label{deriH}
	\frac{d\tilde{H}}{dt}\ket{\tilde{\psi}}=0.
\end{equation}
Considering $\ket{\tilde{\psi}}=\exp\bigl[i\int_{0}^{t}dt\braket{H}\bigr]\ket{\psi}$, it is easy to conclude that $\ket{\tilde{\psi}}$ satisfies the Schr\"{o}dinger equation with the optimal Hamiltonian $\tilde{H}$:
\begin{equation}\label{new_Schro}
	i\ket{\tilde{\psi}} = \tilde{H}\ket{\tilde{\psi}} ~~~ \hbox{and}~~~P=\ket{\psi}\bra{\psi} =\ket{\tilde{\psi}}\bra{\tilde{\psi}}=\tilde{P}.
\end{equation} 
Using eqs.\ (\ref{new_Schro}) and (\ref{H_OP}), we find
\begin{equation}\label{tildeH}
	\tilde{H}=i\biggl(\ket{\dot{\tilde{\psi}}}\bra{\tilde{\psi}}-\ket{\tilde{\psi}}\bra{\dot{\tilde{\psi}}}\biggr).
\end{equation} 
eq.(\ref{avg_H}) implies that $\braket{\dot{\tilde{\psi}}|\tilde{\psi}}=0$. Then using eqs.\ (\ref{tildeH}) and (\ref{new_Schro}), \ (\ref{deriH}) implies
\begin{equation}\label{opti_t}
	(1-\tilde{P})\ket{\ddot{\tilde{\psi}}}=0,
\end{equation}
which is the optimal trajectory for the quantum brachistochrone. Using eqs.\ (\ref{tildeH}), (\ref{opti_t}) and (\ref{avg_H}), we get
\begin{equation}\label{H0}
	\dot{\tilde{H}}=0.
\end{equation}
Now using equation eqs.\ (\ref{new_Schro}) and (\ref{H0}), the differential equation for the optimal trajectory becomes
\begin{equation}\label{eq161}
	\ket{\ddot{\tilde{\psi}}} = -\omega^2
	\ket{\tilde{\psi}},
\end{equation}
where we have used eq.(\ref{const}) $\braket{\tilde{\psi}|\tilde{H}^2|\tilde{\psi}}=(\Delta E)^2 = \omega^2$. One can solve eq.(\ref{eq161}) with the initial conditions on $\ket{\tilde{\psi}(0)}$ and $\ket{\dot{\tilde{\psi}}(0)}$ to get
\begin{equation}\label{op_tr}
	\ket{\tilde{\psi}(t)}=\cos(\omega t) \ket{\tilde{\psi}(0)}+\frac{\sin (\omega t)}{\omega} \ket{\dot{\tilde{\psi}}(0)}
\end{equation}

Thus, the Dirac--Bergmann approach correctly reproduces the known results \cite{carlini2005quantum} for the optimal Hamiltonian (\ref{H_OP}) and the optimal trajectories (\ref{op_tr}).

\subsection{The optimal control of open quantum systems}
The dynamics of an open system under the Markovian approximation is given by the Lindblad master equation \cite{lindblad1976generators,gorini1976completely, lidar2019lecture, breuer2002theory}
\begin{equation}
\dot{\rho} := \mathcal{L}(\rho) = -i[H, \rho] + \sum_a \left( L_a \rho L_a^\dagger - \frac{1}{2} \{ L_a^\dagger L_a, \rho \} \right)
\end{equation}
where $\rho(t)$ is the reduced density matrix of the system, $H(t)$ is the Hamiltonian, $L_a(t);~~a=1,2,...,(N^2-1)$ are the Lindblad operators and $N$ is the dimension of the system Hilbert space. 

Following Koike {\it et al} \cite{carlini2007quantum}, we use the gauge invariance of the Lindblad equation to render the Hamiltonian and the Lindblad operators traceless:
\begin{align}
    \mathrm{Tr}(H)\to0;~~~\mathrm{Tr}(L)\to0.
\end{align}
Also, using the freedom of the bath basis, we could consider the Lindblad operators to be mutually orthonormal:
\begin{align}
    \mathrm{Tr}(L^{\dagger}_{a}L_b)=N \sum_a\gamma_a^2 \delta_{ab}.
\end{align}
The Lagrangian for the time-optimal control problem of the open system is given by \cite{carlini2007quantum}
\begin{equation}\label{l_o_oqs}
    \mathbf{L} = 1 + \operatorname{Tr}[\sigma(\dot{\rho} - \mathcal{L}(\rho))] +  \lambda f(H) + \sum_{a,b} \mu_{ab}[\operatorname{Tr}(L_a^\dagger L_b) - N\gamma_a^2 \delta_{ab}],
\end{equation}
where $\sigma$ is a traceless Hermitian operator which enforces the constraint arising from the Lindbladian dynamics $\dot{\rho}(t)=\mathcal{L}(\rho)$. $\mu_{ab}=\mu^{\star}_{ba}$ are complex functions enforcing the Lindblad operators to be mutually orthonormal. $\lambda$ is a real function that enforces any energy constraint $f(H)=0$. Note that here we only consider one energy constraint, but it can be easily generalized to more than one constraint without loss of generality.

The canonically conjugate momenta are
\begin{align}
\nonumber
P^\rho = \frac{\partial \mathbf{L}}{\partial \dot{\rho}}=\sigma ;~~~P^\sigma = \frac{\partial \mathbf{L}}{\partial \dot{\sigma}}=0 ;~~~P^H = \frac{\partial \mathbf{L}}{\partial \dot{H}}=0 ;~~~P^\lambda = \frac{\partial \mathbf{L}}{\partial \dot{\lambda}}=0 ;\\
P^\mu_{ab} = \frac{\partial \mathbf{L}}{\partial \dot{\mu}_{ab}}=0 ;~~~P^{L}_a = \frac{\partial \mathbf{L}}{\partial \dot{L}_a}=0 ;~~~P^{L^{\dagger}}_a = \frac{\partial \mathbf{L}}{\partial \dot{L^{\dagger}_a}}=0,
\end{align}
with the following Poisson brackets:
\begin{align}
\nonumber
    \left\{\rho, P^\rho\right\}=1;~~~\left\{\sigma,P^\sigma\right\}=1;~~~\left\{H,P^H\right\}=1;~~~\left\{\lambda, P^\lambda\right\}=1;\\
    \left\{\mu_{ab}, P^\mu_{cd}\right\}=\delta_{ac}\delta_{bd};~~~\left\{L_a, P^L_b\right\}=\delta_{ab};~~~\left\{L^{\dagger}_a, P^{L^{\dagger}}_b\right\}=\delta_{ab}.
\end{align}
The primary constraints are
\begin{align}
\label{poqs1}
\chi^{\rho} = P^{\rho}-\sigma \approx 0;~~~\chi^{\sigma} = P^{\sigma} \approx 0;~~~\chi^{H} = P^{H} \approx 0;~~~\chi^{\lambda} = P^{\lambda} \approx 0;\\
\label{poqs2}
\chi^{\mu}_{ab} = P^{\mu}_{ab} \approx 0;~~~\chi^{L}_a = P^{L}_a \approx 0;~~~\chi^{L^{\dagger}}_a = P^{L^{\dagger}}_a \approx 0.
\end{align}
The canonical Hamiltonian is
\begin{align}
\nonumber
    H_c = \operatorname{Tr}\left[P^\rho \dot{\rho}+P^\sigma \dot{\sigma} +P^H \dot{H}+P^\lambda \dot{\lambda}+\sum_{a,b}P^\mu_{ab} \dot{\mu}_{ab}+\sum_{a}(P^L_a \dot{L}_a+P^{L^\dagger}_a \dot{L}^{\dagger}_a)\right]-\mathbf{L} \\
    =-1+\operatorname{Tr}(\sigma \mathcal{L}(\rho))-\lambda f(H) - \sum_{a,b} \mu_{ab} \left[\operatorname{Tr}\left(L^{\dagger}_{a}L_b\right)-N \gamma^{2}_{a}\delta_{ab}\right],
\end{align}
whereby the primary Hamiltonian is seen to be
\begin{align}
\nonumber
    H_p = 
    -1+\operatorname{Tr}(\sigma \mathcal{L}(\rho))-\lambda f(H) - \sum_{a,b} \mu_{ab} \left[\operatorname{Tr}\left(L^{\dagger}_{a}L_b\right)-N \gamma^{2}_{a}\delta_{ab}\right] \\
    + \Lambda^\rho (P^\rho - \sigma) + \Lambda^\sigma P^\sigma + \Lambda^H P^H + \Lambda^\lambda P^\lambda +\sum_{a,b} \Lambda^\mu_{ab} P^\mu_{ab} + \sum_a(\Lambda^L_a P^L_a + \Lambda^{L^{\dagger}}_a P^{L^{\dagger}}_a),
\end{align}
where $\Lambda^\rho$, $\Lambda^\sigma$, $\Lambda^H$, $\Lambda^\lambda$, $\Lambda^{\mu}$, $\Lambda^L$ and $\Lambda^{L^\dagger}$ are other Lagrange multipliers which incorporates the primary constraints $\chi^{\rho}$, $\chi^{\sigma}$, $\chi^{H}$,  $\chi^{\lambda}$, $\chi^\mu$, $\chi^{L}$ and $\chi^{L^\dagger}$  respectively in to the primary Hamiltonian.

We check the consistency of the primary constraints (\ref{poqs1}), (\ref{poqs2}) as follows from eq.(\ref{consistency}) :
\begin{align}
&\dot{\chi}^\rho 
    = \mathcal{L}^\dagger(\sigma) - \Lambda^{\sigma} \neq 0 
    \implies \Lambda^\sigma = \mathcal{L}^\dagger(\sigma)
\\[6pt]
&\dot{\chi}^\sigma 
    = -\mathcal{L}(\rho) + \Lambda^{\rho} \neq 0 
    \implies \Lambda^{\rho} = \mathcal{L}(\rho),
\end{align}
which fixes $\Lambda^\sigma$ and $\Lambda^\rho$ and gives us no secondary constraints. Similarly,
\begin{align}
\nonumber
\dot{\chi}^H 
    = i[\rho, \sigma] + \lambda f^{\prime}(H) \neq 0
\end{align}
gives us a secondary constraint
\begin{align}
    \Sigma^H =  i[\rho, \sigma]+\lambda f^{\prime}(H) \approx 0.
\end{align}
Working through the rest of the variables in the same manner, we obtain the following secondary constraints:
\begin{align}
\nonumber\dot{\chi}^\lambda 
&= f(H) \neq 0 \\
\implies \Sigma^{\lambda} &= f(H) \approx 0;\\
\nonumber
\dot{\chi}^{\mu}_{cd} 
    &= \operatorname{Tr}\bigl(L^{\dagger}_{c} L_{d}\bigr) 
    - N \gamma^{2}_{a} \delta_{cd} \neq 0 \\[6pt]
\implies \Sigma^{\mu}_{cd} 
    &= \operatorname{Tr}\bigl(L^{\dagger}_{c} L_{d}\bigr) 
    - N \gamma^{2}_{a} \delta_{cd} \approx 0;\\
\nonumber
\dot{\chi}_{c}^{L} 
\nonumber
&= -\rho L^{\dagger}_{c} \sigma 
    + \tfrac{1}{2} \bigl\{ \rho, \sigma \bigr\} L^{\dagger}_{c} 
    + \sum_{a} \mu_{ac} L^{\dagger}_{a} \neq 0 \\[6pt]
\implies 
\Sigma^{L}_{c} 
&= -\rho L^{\dagger}_{c} \sigma 
    + \tfrac{1}{2} \bigl\{ \rho, \sigma \bigr\} L^{\dagger}_{c} 
    + \sum_{a} \mu_{ac} L^{\dagger}_{a} 
    \approx 0;\\
\nonumber
\dot{\chi}_{c}^{L^\dagger} 
\nonumber
&= -\sigma L_{c} \rho 
    + \tfrac{1}{2} L_{c} \bigl\{ \rho, \sigma \bigr\}
    + \sum_{a} \mu_{ca} L_{a} \neq 0 \\[6pt]
\implies 
\Sigma^{L^\dagger}_{c} 
&= -\sigma L_{c} \rho 
    + \tfrac{1}{2} L_{c} \bigl\{ \rho, \sigma \bigr\}
    + \sum_{a} \mu_{ca} L_{a} 
    \approx 0.
\end{align}
Now we have to check the consistency of the secondary constraints $\Sigma^H$, $\Sigma^\lambda$, $\Sigma^\mu$, $\Sigma^{L}$, $\Sigma^{L^{\dagger}}$:
\begin{align}\label{sigma_lam}
    \dot{\Sigma}^\lambda=\Lambda^H f^{\prime}(H) \neq 0 \implies \Lambda^H =0, ~~ \text{since}~ f^{\prime}\neq 0;
\end{align}
\begin{align}\label{sigma_h}
    \dot{\Sigma}^H = \Lambda^\lambda f^{\prime}(H)+\Lambda^H \lambda f^{\prime\prime}(H)+2i(\Lambda^\sigma \rho +\Lambda^\rho \sigma)\neq 0.
\end{align}
From eq.(\ref{sigma_lam}) we see that the above eq.(\ref{sigma_h}) implies that 
\begin{align}
    \Lambda^\lambda = -2i \frac{\Lambda^\rho \sigma + \Lambda^\sigma \rho}{f^{\prime}(H)}.
\end{align}
Since the Lagrange multipliers $\Lambda^{\lambda}$ and $\Lambda^{H}$ fix the consistency of $\Sigma_H$ and $\Sigma^\lambda$, there won't be any more secondary constraints generated from them. Further
\begin{align}
\nonumber
\dot{\Sigma}^{\mu}_{cd} 
= \Lambda^{L}_{d} L^{\dagger}_{c} 
    + \Lambda^{L^\dagger}_{c} L_{d} 
    \neq 0.
\end{align}
Since $L^{\dagger}_c$ and $L_d$ are non-zero and linearly independent, we have
\begin{align}
    \Lambda^{L}_{d}=0; ~~~ \Lambda^{L^\dagger}_{c}=0~~~ \forall c,d.
\end{align}
Thus $\Sigma^{\mu}_{cd}$ won't generate further secondary constraints. As for $L$ and $L^\dagger$,
\begin{align}
\label{idn1}
\dot{\Sigma}^{L}_{c}
&= \tfrac{1}{2} [\,\sigma, L^{\dagger}_{c}\,] \Lambda^\rho 
    + \sum_{a} L^{\dagger}_{a} \Lambda^{\mu}_{ac} +\sum_a \mu_{ac} \Lambda^{L^{\dagger}}_{a}+\frac{1}{2}[\sigma,\rho]\Lambda^{L^\dagger}_{c}
    \neq 0\\
    &\label{last1} \implies 
    \sum_{a} L^{\dagger}_{a} \Lambda^{\mu}_{ac} 
     = \tfrac{1}{2} [\,L^{\dagger}_{c}, \sigma\,] \Lambda^\rho;\\
\label{idn2}
\dot{\Sigma}^{L^\dagger}_{c}
&= [\,L_{c}, \sigma\,] \Lambda^\rho 
    + \sum_{b} L_{b} \Lambda^{\mu}_{cb} + \frac{1}{2}[\rho,\sigma]\Lambda^{L}_{c}+\sum_a \mu_{ca}\Lambda^{L}_{a}
    \neq 0\\
    \label{last2}&\implies 
    \sum_{b} L_{b} \Lambda^{\mu}_{cb} 
    = [\,\sigma, L_{c}\,] \Lambda^\rho,
\end{align}
where we have used $\Lambda^{L}_{a}=0$ and $\Lambda^{L^\dagger}_{a}=0~~\forall a$ in eqs.(\ref{idn1}) and(\ref{idn2}) to get eqs.(\ref{last1}) and (\ref{last2}).

The above (\ref{last1}) and (\ref{last2}) are in total six equations for the $\Lambda^\mu_{ab}$. Since there are only six independent $\Lambda^\mu$ due to the fact that $\mu_{ab}=\mu^{\star}_{ba}$, we can, in principle, solve for all the six Lagrange multipliers related to $\Lambda^\mu$. Thus there won't be further secondary constraints generated from $\Sigma^L$ and $\Sigma^{L^{\dagger}}$.

Thus all the primary constraints are preserved over time via generating some new secondary constraints along with solving for the Lagrangian multipliers. All the secondary constraints are relevant for finding the equations for the optimal control; we list them below.
\begin{align}
    \label{sigma1}
    &\Sigma^H =  i[\rho, \sigma]+\lambda f^{\prime}(H) \approx 0\\
    &\Sigma^{\lambda} = f(H) \approx 0\\
    &\Sigma^{\mu}_{cd}=\operatorname{Tr}(L^{\dagger}_{c}L_{d})-N\gamma^{2}_{a}\delta_{cd}\approx 0\\
    &\Sigma^{L}_{c}=-\rho L^{\dagger}_{c}\sigma + \frac{1}{2} \{\rho, \sigma\}L^{\dagger}_{c} +\sum_{a}\mu_{ac}L^{\dagger}_{a}\approx 0\\
    \label{sigma5}
    &\Sigma^{L^\dagger}_{c}=-\sigma L_{c}\rho + \frac{1}{2} L_{c} \{\rho, \sigma\}+\sum_{a}\mu_{ca}L_{a}\approx 0
\end{align}

One gets the same set of equations (\ref{sigma1})--(\ref{sigma5}) by the usual variational approach to an optimal control problem, i.e.,\ varying the action corresponding to the Lagrangian (\ref{l_o_oqs}) of the optimal problem of an open quantum system. As emphasized earlier, these equations follow algebraically in the Dirac--Bergmann approach from the closure of the Poisson Bracket algebra of the full set of constraints. From here on, the specifics of any given optimal control problem can be combined with the above general conditions to solve for the optimal controls and trajectories.

\section{Summary}\label{S6}
In this paper, we developed a formalism that enables us to use the elegant Dirac--Bergmann algorithm, originally developed to study constrained dynamical systems, in the context of optimal control theory. It is shown that, compared to the Pontryagin maximum principle---usually used to study optimal control problems---the Dirac--Bergmann method has some advantages. In particular, because the constraints are required at each stage to be preserved over time, the Dirac--Bergmann method automatically picks out the optimal controls when we restrict the dynamics to the physical subspace. As a spin-off, we can further easily identify the optimal trajectories by solving the relevant equations of motion involving Dirac brackets. We have illustrated this explicitly by considering the example of the brachistochrone, a prototypical example of an optimization problem, both in the classical and quantum cases.


\section{Future directions}
\label{S7}

This paper brings to bear the power of the Dirac--Bergmann algorithm on optimal control theory. The convergence of two different methods developed with diverse interests opens up several hitherto unexplored avenues. This paper is the first of a series of papers in which we plan to explore the following themes: 
\begin{enumerate}
\item  Democratizing optimal control theory: Standard methods of control theory incorporate the equations of motion of the system through dynamical constraints, while the controls are usually treated as parameters which 
can be tuned at will to accomplish the desiderata. However, it is not unreasonable to expect the controls themselves to be governed by some equations of motion. By elevating the controls to be dynamical, we would treat the control and the controlled on an equal footing. We are currently exploring the ramifications of this generalization---in particular, its connection to the strong PMP and the so-called \emph{bang-bang} solutions for the optimal controls, which were not considered in this paper. We expect a ``dynamicalization'' of the controls to naturally yield approximations to such solutions, without the need for global optimization techniques.

\item Fully quantum control theory: In the Copenhagen interpretation of quantum mechanics, the way to make measurements on a quantum system is to couple it to an apparatus axiomatically labeled ``classical'' \cite{LandauLifshitz1977}. A fully quantum model of measurement was later developed by von Neumann \cite {vonNeumann1955}. In a similar spirit, it would be interesting to develop a fully quantum optimal control theory, as opposed to current hybrid theories, which use classical optimization on quantum systems, as we did in this paper. 

\item  Non-local controls: It is conceivable that optimizing complex systems will require multiple controls, each with its own dynamics, operating in an ordered sequence. In particular this might require the coupling between the system and the controls to be non-local, or the controls themselves to be non-local. We are currently exploring ways to construct and study such
models. 
\end{enumerate}

Besides the aforementioned avenues, some other ideas follow as natural extensions to the present work, viz.\ applying optimal control theory to the relativistic, fermionic, and quantum field-theoretic brachistochrones. In the special case where the field theories have gauge invariance, the Dirac--Bergmann approach developed in this paper may be particularly well-suited.  
      
In the present paper, we needed to compute the inverse of a 12$\times$12 matrix in the classical case, and an 8$\times$8 matrix in the quantum case, to implement the Dirac--Bergmann algorithm. It is to be expected that as the systems we try to optimize become larger and more complex, the inverse of the matrix of Poisson brackets that appears in the Dirac--Bergmann algorithm becomes more and more cumbersome to compute. It is desirable to develop an algorithm that automates such computations using software tools. 

On that note, algorithms such as GRAPE, CRAB, and Krotov's algorithm have shown limited success in the implementation of optimal control for open quantum systems \cite{breuer2002theory,carmichael2013statistical}.  Under the Born--Markov approximation, there has been some steady progress over the last few decades; however, optimal control of open quantum systems still remains a challenging problem, not to mention that we know very little about the control of non-Markovian \cite{breuer2002theory} quantum systems. Our proof-of-principle foray into open-system quantum control in the current work serves as a small step towards tackling the more general challenge. Two of the authors ~\cite{maity2024dirac} have earlier explored connections between the Dirac--Bergmann algorithm and Lindblad equations. It would be interesting to understand the connections between the three threads in a more holistic fashion, with the potential to develop open system control algorithms that overcome the limitations of the existing ones.

\section*{Acknowledgments}
DA would like to thank Armen Nersessian for introducing the Dirac--Bergman algorithm to him, and Foo Yan Xi for fruitful discussions. This work is partially funded by a grant from the Infosys Foundation.

\bibliographystyle{unsrt}
\bibliography{Dirac_control}

\appendix
\section*{Appendix}
The matrix of Poisson brackets of all the second-class constraints is given by eqs.(\ref{cmn}) and  (\ref{sc1})-(\ref{CL12})

\[D_{\alpha\beta}=
\left(
\scalebox{0.7}{$
	\begin{array}{cccccccccccc}
		0 & 0 & 0 & 0 & -1 & 0 & 0 & 0 & 0 & 0 & 0 & 0 \\
		0 & 0 & 0 & 0 & 0 & -1 & 0 & 0 & 0 & -\frac{\Lambda_4}{x_4} & \frac{x_4 \Lambda_3 + g \Lambda_4}{x_4^2} & \frac{2g x_4 (-x_2^2 + x_4^2)}{(x_2^2 + x_4^2)^2} \\
		0 & 0 & 0 & 0 & 0 & 0 & -1 & 0 & 0 & 0 & 0 & 0 \\
		0 & 0 & 0 & 0 & 0 & 0 & 0 & -1 & 0 & \frac{x_2 \Lambda_4}{x_4^2} & -\frac{x_2 (x_4 \Lambda_3 + 2g \Lambda_4)}{x_4^3} & \frac{2g x_2 (x_2^2 - x_4^2)}{(x_2^2 + x_4^2)^2} \\
		1 & 0 & 0 & 0 & 0 & 0 & 0 & 0 & 0 & 0 & -1 & 0 \\
		0 & 1 & 0 & 0 & 0 & 0 & 0 & 0 & 0 & 1 & 0 & 0 \\
		0 & 0 & 1 & 0 & 0 & 0 & 0 & 0 & 0 & 0 & \frac{x_2}{x_4} & 0 \\
		0 & 0 & 0 & 1 & 0 & 0 & 0 & 0 & 0 & -\frac{x_2}{x_4} & \frac{g x_2}{x_4^2} & 0 \\
		0 & 0 & 0 & 0 & 0 & 0 & 0 & 0 & 0 & 0 & 0 & -1 \\
		0 & \frac{\Lambda_4}{x_4} & 0 & -\frac{x_2 \Lambda_4}{x_4^2} & 0 & -1 & 0 & \frac{x_2}{x_4} & 0 & 0 & 0 & 0 \\
		0 & -\frac{x_4 \Lambda_3 + g \Lambda_4}{x_4^2} & 0 & \frac{x_2 (x_4 \Lambda_3 + 2g \Lambda_4)}{x_4^3} & 1 & 0 & -\frac{x_2}{x_4} & -\frac{g x_2}{x_4^2} & 0 & 0 & 0 & 0 \\
		0 & \frac{2g x_4 (x_2^2 - x_4^2)}{(x_2^2 + x_4^2)^2} & 0 & -\frac{2g x_2 (x_2^2 - x_4^2)}{(x_2^2 + x_4^2)^2} & 0 & 0 & 0 & 0 & 1 & 0 & 0 & 0
	\end{array}
	$}
\right).
\]
\\
The inverse of this matrix is too large to display in full within the page. Thus, its non-zero components are given explicitly as
\begin{align*}
	(D^{-1})_{1,2} = \frac{x_4^4}{(x_2^2 + x_4^2)(x_4 \Lambda_3 + g \Lambda_4)} 
	\qquad \qquad (D^{-1})_{1,4} = -\frac{x_2 x_4^3}{(x_2^2 + x_4^2)(x_4 \Lambda_3 + g \Lambda_4)}
\end{align*}
\vspace{-15pt}
\begin{align*}
	(D^{-1})_{1,6} &= 1 \
	&(D^{-1})_{1,7} = \frac{x_4^3 \Lambda_4}{(x_2^2 + x_4^2)(x_4 \Lambda_3 + g \Lambda_4)}
\end{align*}
\vspace{-15pt}
\begin{align*}
	(D^{-1})_{1,9} &= -\frac{x_2 x_4^2 \Lambda_4}{(x_2^2 + x_4^2)(x_4 \Lambda_3 + g \Lambda_4)} \
	&(D^{-1})_{1,10} = \frac{2 g x_4^3 (-x_2^2 + x_4^2)}{(x_2^2 + x_4^2)^2 (x_4 \Lambda_3 + g \Lambda_4)}
\end{align*}
\vspace{-15pt}
\begin{align*}
	(D^{-1})_{2,1} &= -\frac{x_4^4}{(x_2^2 + x_4^2)(x_4 \Lambda_3 + g \Lambda_4)} \
	&(D^{-1})_{2,3} = \frac{x_2 x_4^3}{(x_2^2 + x_4^2)(x_4 \Lambda_3 + g \Lambda_4)}
\end{align*}
\vspace{-15pt}
\begin{align*}
	(D^{-1})_{2,5} &= \frac{g x_2 x_4^2}{(x_2^2 + x_4^2)(x_4 \Lambda_3 + g \Lambda_4)} \
	&(D^{-1})_{2,7} = \frac{x_2^2}{x_2^2 + x_4^2}
\end{align*}
\vspace{-15pt}
\begin{align*}
	(D^{-1})_{2,9} &= \frac{x_2 x_4 (x_4 \Lambda_3 + 2 g \Lambda_4)}{(x_2^2 + x_4^2)(x_4 \Lambda_3 + g \Lambda_4)} \
	&(D^{-1})_{2,10} = \frac{2 g^2 x_2^2 x_4^2 (x_2^2 - x_4^2)}{(x_2^2 + x_4^2)^3 (x_4 \Lambda_3 + g \Lambda_4)}
\end{align*}
\vspace{-15pt}
\begin{align*}
	(D^{-1})_{3,2} &= -\frac{x_2 x_4^3}{(x_2^2 + x_4^2)(x_4 \Lambda_3 + g \Lambda_4)} \
	&(D^{-1})_{3,4} = \frac{x_2^2 x_4^2}{(x_2^2 + x_4^2)(x_4 \Lambda_3 + g \Lambda_4)}
\end{align*}
\vspace{-15pt}
\begin{align*}
	(D^{-1})_{3,7} &= -\frac{x_2 x_4^2 \Lambda_4}{(x_2^2 + x_4^2)(x_4 \Lambda_3 + g \Lambda_4)} \
	&(D^{-1})_{3,8} = 1
\end{align*}
\vspace{-15pt}
\begin{align*}
	(D^{-1})_{3,9} &= \frac{x_2^2 x_4 \Lambda_4}{(x_2^2 + x_4^2)(x_4 \Lambda_3 + g \Lambda_4)} \
	&(D^{-1})_{3,10} = \frac{2 g x_2 x_4^2 (x_2^2 - x_4^2)}{(x_2^2 + x_4^2)^2 (x_4 \Lambda_3 + g \Lambda_4)}
\end{align*}
\vspace{-15pt}
\begin{align*}
	(D^{-1})_{4,1} &= \frac{x_2 x_4^3}{(x_2^2 + x_4^2)(x_4 \Lambda_3 + g \Lambda_4)} \
	&(D^{-1})_{4,2} = -\frac{g x_2 x_4^2}{(x_2^2 + x_4^2)(x_4 \Lambda_3 + g \Lambda_4)}
\end{align*}
\vspace{-15pt}
\begin{align*}
	(D^{-1})_{4,3} &= -\frac{x_2^2 x_4^2}{(x_2^2 + x_4^2)(x_4 \Lambda_3 + g \Lambda_4)} \
	&(D^{-1})_{4,7} = \frac{x_2 x_4^2 \Lambda_3}{(x_2^2 + x_4^2)(x_4 \Lambda_3 + g \Lambda_4)}
\end{align*}
\vspace{-15pt}
\begin{align*}
	(D^{-1})_{4,9} &= \frac{x_4^2}{x_2^2 + x_4^2} \
	&(D^{-1})_{4,10} = \frac{2 g^2 x_2 x_4^3 (x_2^2 - x_4^2)}{(x_2^2 + x_4^2)^3 (x_4 \Lambda_3 + g \Lambda_4)}
\end{align*}
\vspace{-15pt}
\begin{align*}
	(D^{-1})_{5,1} &= -1 \
	&(D^{-1})_{6,2} = -\frac{x_4^3 \Lambda_4}{(x_2^2 + x_4^2)(x_4 \Lambda_3 + g \Lambda_4)}
\end{align*}
\vspace{-15pt}
\begin{align*}
	(D^{-1})_{6,3} &= \frac{x_2 x_4^2 \Lambda_4}{(x_2^2 + x_4^2)(x_4 \Lambda_3 + g \Lambda_4)} \
	&(D^{-1})_{6,4} = -\frac{x_2 x_4^2 \Lambda_3}{(x_2^2 + x_4^2)(x_4 \Lambda_3 + g \Lambda_4)}
\end{align*}
\vspace{-15pt}
\begin{align*}
	(D^{-1})_{6,9} &= \frac{g x_2 \Lambda_4^2}{(x_2^2 + x_4^2)(x_4 \Lambda_3 + g \Lambda_4)} \
	&(D^{-1})_{6,10} = \frac{2 g^2 x_2^2 x_4 (x_2^2 - x_4^2) \Lambda_4}{(x_2^2 + x_4^2)^3 (x_4 \Lambda_3 + g \Lambda_4)}
\end{align*}
\vspace{-15pt}
\begin{align*}
	(D^{-1})_{6,11} &= -\frac{x_4^2}{x_2^2 + x_4^2} \
	&(D^{-1})_{6,12} = -\frac{x_4^3 \Lambda_4}{(x_2^2 + x_4^2)(x_4 \Lambda_3 + g \Lambda_4)}
\end{align*}
\vspace{-15pt}
\begin{align*}
	(D^{-1})_{7,3} &= -1 \
	&(D^{-1})_{8,1} = \frac{x_2 x_4^2 \Lambda_4}{(x_2^2 + x_4^2)(x_4 \Lambda_3 + g \Lambda_4)}
\end{align*}
\vspace{-15pt}
\begin{align*}
	(D^{-1})_{8,2} &= -\frac{x_2 x_4 (x_4 \Lambda_3 + 2 g \Lambda_4)}{(x_2^2 + x_4^2)(x_4 \Lambda_3 + g \Lambda_4)} \
	&(D^{-1})_{8,3} = -\frac{x_2^2 x_4 \Lambda_4}{(x_2^2 + x_4^2)(x_4 \Lambda_3 + g \Lambda_4)}
\end{align*}
\vspace{-15pt}
\begin{align*}
	(D^{-1})_{8,4} &= -\frac{x_4^2}{x_2^2 + x_4^2} \
	&(D^{-1})_{8,7} = -\frac{g x_2 \Lambda_4^2}{(x_2^2 + x_4^2)(x_4 \Lambda_3 + g \Lambda_4)}
\end{align*}
\vspace{-15pt}
\begin{align*}
	(D^{-1})_{8,10} &= \frac{2 g^2 x_2 x_4^2 (x_2^2 - x_4^2) \Lambda_4}{(x_2^2 + x_4^2)^3 (x_4 \Lambda_3 + g \Lambda_4)} \
	&(D^{-1})_{9,1} = \frac{2 g x_4^3 (x_2^2 - x_4^2)}{(x_2^2 + x_4^2)^2 (x_4 \Lambda_3 + g \Lambda_4)}
\end{align*}
\vspace{-15pt}
\begin{align*}
	(D^{-1})_{9,2} &= -\frac{2 g^2 x_2^2 x_4^2 (x_2^2 - x_4^2)}{(x_2^2 + x_4^2)^3 (x_4 \Lambda_3 + g \Lambda_4)} \
	&(D^{-1})_{9,3} = \frac{2 g x_2 x_4^2 (-x_2^2 + x_4^2)}{(x_2^2 + x_4^2)^2 (x_4 \Lambda_3 + g \Lambda_4)}
\end{align*}
\vspace{-15pt}
\begin{align*}
	(D^{-1})_{9,4} &= \frac{2 g^2 x_2 x_4^3 (-x_2^2 + x_4^2)}{(x_2^2 + x_4^2)^3 (x_4 \Lambda_3 + g \Lambda_4)} \
	&(D^{-1})_{9,7} = -\frac{2 g^2 x_2^2 x_4 (x_2^2 - x_4^2) \Lambda_4}{(x_2^2 + x_4^2)^3 (x_4 \Lambda_3 + g \Lambda_4)}
\end{align*}
\vspace{-15pt}
\begin{align*}
	(D^{-1})_{9,9} &= -\frac{2 g^2 x_2 x_4^2 (x_2^2 - x_4^2) \Lambda_4}{(x_2^2 + x_4^2)^3 (x_4 \Lambda_3 + g \Lambda_4)} \
	&(D^{-1})_{9,10} = \frac{2 g^2 x_2^2 x_4^2 (x_2^2 - x_4^2)}{(x_2^2 + x_4^2)^3 (x_4 \Lambda_3 + g \Lambda_4)}
\end{align*}
\vspace{-15pt}
\begin{align*}
	(D^{-1})_{9,11} &= \frac{2 g x_4^3 (x_2^2 - x_4^2)}{(x_2^2 + x_4^2)^2 (x_4 \Lambda_3 + g \Lambda_4)} \
	&(D^{-1})_{9,12} = 1
\end{align*}
\vspace{-15pt}
\begin{align*}
	(D^{-1})_{10,1} &= \frac{x_4^4}{(x_2^2 + x_4^2)(x_4 \Lambda_3 + g \Lambda_4)} \
	&(D^{-1})_{10,3} = -\frac{x_2 x_4^3}{(x_2^2 + x_4^2)(x_4 \Lambda_3 + g \Lambda_4)}
\end{align*}
\vspace{-15pt}
\begin{align*}
	(D^{-1})_{10,4} &= -\frac{g x_2 x_4^2}{(x_2^2 + x_4^2)(x_4 \Lambda_3 + g \Lambda_4)} \
	&(D^{-1})_{10,7} = \frac{x_4^2}{x_2^2 + x_4^2}
\end{align*}
\vspace{-15pt}
\begin{align*}
	(D^{-1})_{10,9} &= -\frac{x_2 x_4 (x_4 \Lambda_3 + 2 g \Lambda_4)}{(x_2^2 + x_4^2)(x_4 \Lambda_3 + g \Lambda_4)} \
	&(D^{-1})_{10,10} = -\frac{2 g^2 x_2^2 x_4^2 (x_2^2 - x_4^2)}{(x_2^2 + x_4^2)^3 (x_4 \Lambda_3 + g \Lambda_4)}
\end{align*}
\vspace{-15pt}
\begin{align*}
	(D^{-1})_{10,12} &= \frac{x_4^4}{(x_2^2 + x_4^2)(x_4 \Lambda_3 + g \Lambda_4)} \
	&(D^{-1})_{11,2} = \frac{x_4^4}{(x_2^2 + x_4^2)(x_4 \Lambda_3 + g \Lambda_4)}
\end{align*}
\vspace{-15pt}
\begin{align*}
	(D^{-1})_{11,4} &= -\frac{x_2 x_4^3}{(x_2^2 + x_4^2)(x_4 \Lambda_3 + g \Lambda_4)} \
	&(D^{-1})_{11,7} = \frac{x_4^3 \Lambda_4}{(x_2^2 + x_4^2)(x_4 \Lambda_3 + g \Lambda_4)}
\end{align*}
\vspace{-15pt}
\begin{align*}
	(D^{-1})_{11,9} &= -\frac{x_2 x_4^2 \Lambda_4}{(x_2^2 + x_4^2)(x_4 \Lambda_3 + g \Lambda_4)} \
	&(D^{-1})_{11,10} = \frac{2 g x_4^3 (-x_2^2 + x_4^2)}{(x_2^2 + x_4^2)^2 (x_4 \Lambda_3 + g \Lambda_4)}
\end{align*}
\vspace{-15pt}
\begin{align*}
	(D^{-1})_{11,12} &= -\frac{x_4^4}{(x_2^2 + x_4^2)(x_4 \Lambda_3 + g \Lambda_4)} \
	&(D^{-1})_{12,9} = -1
\end{align*}
\vspace{-15pt}


\end{document}